\documentclass{article}

\usepackage{amssymb,amsfonts,amsmath}
\usepackage{cite,enumerate,float,indentfirst}
\usepackage{color}

\def\be{\begin{eqnarray}}
\def\ee{\end{eqnarray}}
\def\nn{\nonumber}

\def\p{\partial}

\definecolor{red}{rgb}{1,0,0}
\definecolor{orange}{rgb}{1,0.5,0}
\definecolor{violet}{rgb}{0.7,0,1}

\def\Schur{{\rm S}}
\def\Mac{{\rm M}}

\def\SchurP{{\rm Ker}}

\def\GMac{{\cal M}}

\def\C{{\cal C}}
\def\CC{C}
\def\CK{{\C}}

\hoffset= - 1.0in         

\begin{document}

\title{\vspace{-.5cm}{\Large {\bf  On generalized Macdonald polynomials}\vspace{.05cm}}
\author{
{\bf A.Mironov$^{a,b,c}$}\footnote{mironov@lpi.ru; mironov@itep.ru}\ \ and
\ {\bf A.Morozov$^{d,b,c}$}\thanks{morozov@itep.ru}}
\date{ }
}

\maketitle

\vspace{-5cm}

\begin{center}
\hfill FIAN/TD-08/19\\
\hfill IITP/TH-11/19\\
\hfill ITEP/TH-18/19\\
\hfill MIPT-TH-09/19
\end{center}

\vspace{3cm}

\begin{center}
$^a$ {\small {\it Lebedev Physics Institute, Moscow 119991, Russia}}\\
$^b$ {\small {\it ITEP, Moscow 117218, Russia}}\\
$^c$ {\small {\it Institute for Information Transmission Problems, Moscow 127994, Russia}}\\
$^d$ {\small {\it MIPT, Dolgoprudny, 141701, Russia}}
\end{center}

\vspace{.5cm}

\begin{abstract}
Generalized Macdonald polynomials (GMP) are eigenfunctions of
specifically-deformed Ruijsenaars Hamiltonians and are
built as triangular polylinear
combinations of Macdonald polynomials.
They are orthogonal with respect to a modified scalar product,
which could be constructed with the help of an increasingly important
triangular perturbation theory, showing up in a variety of applications.
A peculiar feature of GMP is that denominators in this expansion
are fully factorized, which is a consequence of a hidden symmetry
resulting from the special choice of the Hamiltonian deformation.
We introduce also a simplified but deformed version of GMP,
which we call generalized Schur functions. Our basic examples are bilinear in Macdonald polynomials.
\end{abstract}

\vspace{.0cm}

\section{Introduction}

 Macdonald polynomials are getting increasingly important for practical calculations in string theory.
The reason is their role in representation theory of the Ding-Iohara-Miki symmetry,
which underlines dynamics of background brane networks in $6d$ super-Yang-Mills models.
At the same time, their properties remain under-investigated, especially for the purposes of
physical applications.
This paper is one in the series of recent attempts to cure this situation.
This time we concentrate on the subject of {\it generalized} Macdonald polynomials (GMP).

Macdonald polynomials $\Mac_R$ labelled by Young diagrams $R$ are \cite{Mac} graded symmetric polynomials of variables $x_i$, $i=1,\ldots,N$ (in this case we use the notation $\Mac_R(x_i)$), or graded polynomials of time-variables $p_k:=\sum_{i=1}^Nx_i^k$ (in this case we use the notation $\Mac_R\{p_k\}$). In this paper, we will be mostly interested in the second point of view.
They can be defined in many different ways, in particular, within the context of quantum toroidal algebras \cite{DIM} (see also \cite{nDIM} and references therein). There are two other natural definitions: using their triangle structure and orthogonality condition w.r.t. a scalar product, or using a Hamiltonian structure behind them. We will briefly review in s.2 both of these possibilities in order to demonstrate that these definitions are really effective.

An essential feature is that the Gaussian averages of Macdonald polynomials are Macdonald dimensions:
this is a basic property
\be
<{\rm character}>\ = character
\label{charchar}
\ee
of matrix and tensor models,
presumably related to their {\it super}integrability, see \cite{MMchar} and
\cite{otherMTmods} for related references.
However, in application to conformal (Dotsenko-Fateev) matrix models \cite{DF}
relevant \cite{AGTmamo,MMS} to the Nekrasov counting \cite{Nek} and the AGT relations \cite{AGT},
there is a significant ``detail":
Macdonald averages, though nicely factorized
(in this case  (\ref{charchar}) is known as Selberg-Kadell identities),
do not coincide with the Nekrasov functions \cite{MMSS}.
The resolution of this problem is that the Macdonald polynomials should be modified
to the generalized Macdonald polynomials \cite{Awata} in such a way that the Selberg-Kadell formulas are still true,
but the Nekrasov formulas are properly reproduced,
see \cite{MorSmi,otherGM} for details.

Unfortunately,
despite their undisputable importance,
these GMP can be effectively described only using the quantum toroidal algebra behind them \cite{Awata,Ohkubo}. That is, the GMP are defined as common eigenfunctions of a deformed
Calogero-Ruijsenaars Hamiltonian (generalized cut-and-join operators of \cite{MMN}) that is read off from the quantum toroidal algebra \cite{Awata,Ohkubo}.
The goal of the present paper is to look for a definition independent on this algebraic approach.
We discuss the two possibilities: how the deformed Hamiltonian can be independently obtained and how the GMP can be defined
through the Gauss decomposition of an appropriate scalar product.

\section{Macdonald polynomials}

\subsection{Triangular structure}

Let us fix the scalar product on time-variables
\be\label{sp}
\Big< { p}_{\Delta}\Big| { p}_{\Delta'} \Big>_{(q,t)}=
z_\Delta \cdot \delta_{\Delta,\Delta'}\cdot
\left(\prod_{i=1}^{l_\Delta} {\{q^{\delta_i}\}\over\{t^{\delta_i}\}}\right)
\ee
and continue it to any polynomials by linearity. From no on, we use the notation $\{x\}:=x-x^{-1}$.
Here the Young diagram $\Delta=\big[\delta_1\geq\delta_2\geq\ldots\geq \delta_{l_\Delta}\big]$,
and $p^\Delta = \prod_{i=1}^{l_\Delta} p_{\delta_i}$.
The combinatorial factor $z_\Delta$ is best defined in the dual parametrization of the
Young diagram, $\Delta = \big[\ldots,2^{m_2},1^{m_1}\big]$, then
$z_\Delta = \prod_k k^{m_k}\cdot m_k!$.

Then, the ordinary Macdonald polynomials can be defined as a lower triangular combination of the Schur polynomials ${\Schur}_{R}\{p\}$
\be\label{PthroughSchurs}
\Mac_R\{ p\} =  {\Schur}_{R}\{p\}
+ \sum_{{R'< R}} {\cal K}_{{R,R'}}(q,t) \cdot {\Schur}_{R'}\{p\}
\ee
orthogonal w.r.t. this scalar product (\ref{sp}):
\be\label{orc}
\Big<\Mac_R\Big|\Mac_{R'}\Big>=||\Mac_R||^2\cdot\delta_{R,R'}
\ee
The Young diagrams are ordered lexicographically:
\be
R>R'  \ \ {\rm if} \ \ r_1>r_1' \ \ {\rm  or\ if} \ \ r_1=r_1', \ {\rm but} \ r_2>r_2',
\ \ {\rm or\  if} \ \ r_1=r_1' \ {\rm  and} \ r_2=r_2', \ {\rm but} \ r_3>r_3',
\ \ {\rm and\ so\ on}
\label{lexico}
\ee
This ordering is not consistent
with the transposition of Young diagrams:
\be
R>R'\ \text{is not always the same as} \ {R'}^\vee >R^\vee
\label{transpolex}
\ee
The first discrepancy appears at level $|R|=6$, it is the pair
$[3,1,1,1]>[2,2,2]$, for which $[3,1,1,1]^\vee = [4,1,1]>[2,2,2]^\vee=[3,3]$.
However, it does not lead to an ambiguity, since the Kostka-Macdonald coefficients ${\cal K}_{{R,R'}}(q,t)$ for this pair of diagrams, and in all similar cases vanishes \cite{Mac}.

Note that the normalization of $\Mac_R$ is already fixed by  the choice of
unit diagonal coefficient (the first term) in (\ref{PthroughSchurs}):
\be
{\cal K}_{R,R}(q,t)=1
\ee
therefore the norm $||\Mac_R||$ is a deducible quantity.

An important point is that one can unambiguously calculate the Macdonald polynomials using this definition. Hence, we have a good definition that does not appeal to any additional algebraic structures.

\subsection{Ruijsenaars Hamiltonians}

Another possible definition of the Macdonald polynomials goes back to S. Ruijsenaars: they are defined as the system of polynomial eigenfunctions of the Ruijsenaars Hamiltonians \cite{Rui}. In fact, there is another approach to constructing Hamiltonians, which admits easier extensions to the Kerov functions: through skew characters and Young diagrams with a given number of hooks (see \cite{triang}). We will return to it somewhere else, and here use the standard framework of the Ruijsenaars Hamiltonians.

\paragraph{The Ruijsenaars Hamiltonians.}
The Ruijsenaars Hamiltonians are formulated in terms of the symmetric variables, $x_i$ and are associated with the particle coordinates in the integrable Ruijsenaars-Schneider system. One can write down a set of $n$ integrable Ruijsenaars Hamiltonians in these variables as difference operators acting on the functions of $N$ variables $x_i$ as
\be\label{Hamk}
\hat H_kF(x_i)=\sum_{i_1<\ldots<i_k}{\prod_{m=1}^kD(t,x_{i_m})\Delta(x)\over \Delta(x)}\prod_{m=1}^kD(q,x_{i_m})F(x_i)
\ee
where $\Delta(x)=\prod_{i<j}(x_i-x_j)$ is the Vandermonde determinant, and $D(\xi,x_i)$ is the operator of dilation of the variable $x_i$: $x_i\to\xi x_i$. The Macdonald polynomials $\Mac_R$ are eigenfunctions of these Hamiltonians, while the generating function of the eigenvalues
\be
\sum_k\lambda_R^{(k)}z^k=\prod_{i=1}\Big(1+zq^{R_i}t^{n-i}\Big)
\ee

\paragraph{The first Hamiltonian in terms of time-variables.}
The Ruijsenaars Hamiltonians can be rewritten in terms of time-variables. The results reads as follows (we slightly redefine the Ruijsenaars Hamiltonians here to make formulas simpler). Introduce an operator
\be
\hat V_m(z):=\exp\left(\sum_{k>0}{(1-t^{-2mk})p_kz^k\over k}\right)\cdot\exp\left(\sum_{k>0}{t^{2mk}-1\over
t^{2k}-1}{q^{2k}-1\over z^k}
{\partial\over\partial p_k}\right)
\ee
Then, the first Hamiltonian is
\be\label{Ham1}
\hat H_1=\oint_0 {dz\over z}\hat V_1(z)
\ee
with the eigenvalues when acting on the eigenfunction $\Mac_R$
\be\label{1ev}
\lambda^{(1)}_R=1+\{t\}\sum_i{q^{2R_i}-1\over t^{2i-1}}
\ee
In fact, in order to construct the Macdonald polynomials, one does not need to know all the Hamiltonians, it is sufficient to use only the first one: its polynomial solutions are unambiguously determined. The reason is the integrable structures behind the system. Nevertheless, in order to have the complete picture, we construct all higher Hamiltonians now.

\paragraph{Higher Hamiltonians.}
The second Hamiltonian is
\be
\hat H_2=\oint_0 {dz\over z}\hat V_2(z)
-{\{t^2\}\over t^2\{t\}^2}
\oint_0 {dz_1\over z_1}\oint_0 {dz_2\over z_2}{(z_1-z_2)^2\over (z_1-t^{-2}z_2)(t^{-2}z_1-z_2)}
:\hat V_1(z_1)\hat V_1(z_2):
\ee

The next Hamiltonian is
\be
\hat H_3=\oint_0 {dz\over z}\hat V_3(z)-{3\over 2}\ {\{t^3\}\over t^2\{t\}\{t^2\}}\oint_0 {dz_1\over z_1}\oint_0 {dz_2\over z_2}
{(z_1-z_2)(t^{-2}z_1-z_2)\over (z_1-t^{-2}z_2)(t^{-4}z_1-z_2)}:\hat V_2(z_1)\hat V_1(z_2):+\nn\\
+{1\over 2}\ {\{t^3\}\over t^6\{t\}^3}\oint_0 {dz_1\over z_1}\oint_0 {dz_2\over z_2}\oint_0 {dz_3\over z_3}
\prod_{i<j}{(z_i-z_j)^2\over (z_i-t^{-2}z_j)(t^{-2}z_i-z_j)}:\hat V_1(z_1)\hat V_1(z_2)\hat V_1(z_3):
\ee

The general Hamiltonian looks like
\be\label{10}
\hat H_k=(-1)^{k+1}\cdot k\cdot \{t^k\}\cdot\sum_\Delta {\psi_{[1^k]}(\Delta)\over z_\Delta}:\left(\prod_{i=1}^{l_{_\Delta}}
\oint_0{dz_i\over z_i}{\hat V_{\delta_i}(z_i)\over t^{2(i-1)}\{t^{\delta_i}\}}\right):
\prod_{i<j}{(z_i-z_j)(t^{-2\delta_i+2}z_i-t^{-2\delta_j+2}z_j)\over (z_i-t^{-2\delta_j}z_j)(t^{-2\delta_i}z_i-z_j)}
\ee
Symbolically, it can be rewritten as
\be\label{11}
\hat H_k=(-1)^{k+1}\cdot k\cdot \{t^k\}\cdot :S_{[1^k]}\Biggr\{p_i=\oint_0{dz_i\over z_i}
{\hat V_{\delta_i}(z_i)\over t^{2(i-1)}\{t^{\delta_i}\}}
\Biggr\}:\cdot
\prod_{i<j}{(z_i-z_j)(t^{-2\delta_i+2}z_i-t^{-2\delta_j+2}z_j)\over (z_i-t^{-2\delta_j}z_j)(t^{-2\delta_i}z_i-z_j)}
\ee
This expression becomes non-symbolic in the case of $q=t$: in this case, the pair product in this expression is a result of normal ordering so that one can re-write
\be
\hat H_k\Big|_{q=t}=(-1)^{k+1}\cdot k\cdot \{t^k\}\cdot  S_{[1^k]}\Biggr\{p_i=\oint_0{dz\over z}
{\hat V_i(z)\over t^{2(i-1)}\{t^i\}}
\Biggr\}
\ee
This is a reflection of the fact that any $\oint_0 dz/z\cdot \hat V_k(z)$ is a Hamiltonian. This is not surprising, since the eigenfunctions in this case do not depend on $t$: these are the Schur functions. Hence, $t$ is a free parameter and such is $t^k$, expansion in any of these parameters generate the corresponding Hamiltonians having the Schur functions as their eigenfunctions.

Unfortunately, in the general case of different $q$ and $t$, the factor coming from the normal ordering is different from that in (\ref{10})-(\ref{11}). For instance,
\be
\prod_i \hat V_1(z_i)=\prod_{i<j}{(z_i-z_j)(q^2z_i-t^2z_j)\over (q^2z_i-z_j)(z_i-t^2z_j)}\ :\prod_i \hat V_1(z_i):
\ee

\paragraph{Eigenvalues.} One would better choose another normalization of the Hamiltonians:
\be
\hat H_k=\sum_\Delta {\psi_{[1^k]}(\Delta)\over z_\Delta}:\left(\prod_{i=1}^{l_{_\Delta}}
\oint_0{dz_i\over z_i}{V_{\delta_i}(z_i)\over t^{2(i-1)}\{t^{\delta_i}\}}\right):
\prod_{i<j}{(z_i-z_j)(t^{-2\delta_i+2}z_i-t^{-2\delta_j+2}z_j)\over (z_i-t^{-2\delta_j}z_j)(t^{-2\delta_i}z_i-z_j)}
\ee
The eigenvalues in symmetric representations are:
\be
\lambda_{[n]}^{(k)}={q^{2n}(t^{2k}-1)+1\over t^{k(k+3)/2}\prod_{i=1}^{k}\{t^i\}}
\ee
Similarly, in representation $[n_1,n_2]$ the eigenvalues are
\be
\lambda_{[n_1,n_2]}^{(k)}={q^{2n_1+2n_2}(t^{2k}-1)(t^{2(k-1)}-1)t^2
+q^{2n_1}(t^{2k}-1)t^2+q^{2n_2}(t^{2k}-1)+1
\over t^{k(k+7)/2}\prod_{i=1}^{k}\{t^i\}}=\nn\\
={Q_1Q_2(t^{2k}-1)(t^{2(k-1)}-1)
+Q_1(t^{2k}-1)+Q_2(t^{2k}-1)+1
\over t^{k(k+7)/2}\prod_{i=1}^{k}\{t^i\}}
\ee
where
\be
Q_i:=q^{2n_i}t^{4-2i}
\ee
In the general case,
\be\label{ev1}
\lambda_{R}^{(k)}={1
\over t^{k(k+4l_{_R}-1)/2}\prod_{i=1}^{k}\{t^i\}}
\left(1+(t^{2k}-1)\sum_iQ_i^{(R)}+(t^{2k}-1)(t^{2k-2}-1)\sum_{i\ne j}Q_i^{(R)}Q_j^{(R)}\right.+\nn\\
+\left.(t^{2k}-1)(t^{2k-2}-1)(t^{2k-4}-1)\sum_{i\ne j\ne l}Q_i^{(R)}Q_j^{(R)}Q_l^{(R)}+\ldots\right)=\nn\\
={1
\over t^{k(k+4l_{_R}-1)/2}\prod_{i=1}^{k}\{t^i\}}\sum_{m=0}^{l_{_R}}\Biggr(\prod_{j=1}^m\Big(t^{2(k-j+1)}-1\Big)
\Biggr)\ \Schur_{[1^m]}\Big(Q_i^{(R)}\Big)=\nn\\
={1
\over t^{k(k+4l_{_R}-1)/2}\prod_{i=1}^{k}\{t^i\}}\sum_{m=0}^{\infty}\Biggr(\prod_{j=1}^m\Big(t^{2(k-j+1)}-1\Big)
\Biggr)\ \Schur_{[1^m]}\Big(Q_i^{(R)}\Big)
\ee
with the symmetric variables
\be\label{sv1}
Q_i^{(R)}:=q^{2R_i}t^{2l_{_R}-2i}
\ee
and the time-variables
\be
p_m^{(R)}:=\sum_i \Big(Q_i^{(R)}\Big)^m
\ee

\paragraph{Another construction of Hamiltonians.}
There is another  construction of the Hamiltonians that can be conveniently extended to the GMP. We borrow the formulas of this paragraph from \cite{MMZ}.

Let us define an operator
\be
\hat P_m:=\oint_{z=0}{dz\over z^{m+1}}\hat V_1(z)
\ee
Then, the new first Hamiltonian coincides with the old one and is equal just to $\hat {\cal H}_1\sim\hat H_1=\hat P_0$, the second one is the commutator $\hat {\cal H}_2\sim [P_{-1},P_1]$ and all remaining Hamiltonians are repeated commutators\footnote{It can be understood within the context of the quantum toroidal algebra, since according to \cite{DIM}, these Hamiltonians are spectral dual (see also \cite{specdu}) to $\psi_k^+$ generators of the algebra and, hence, are commuting.}
\be\label{Ham2}
\hat {\cal H}_k={g_1\over\{q\}^k\{t^{-1}\}^k}[[\ldots[\hat P_{-1},\underbrace{\hat P_0],\ldots,\hat P_0}_{k-2\ times}],\hat P_1]
\ee
where we introduced the notation $g_k:=\{q^k\}\{t^{-k}\}\{t^k/q^k\}$ and a non-unit pre-factor is chosen to simplify further formulas.
The eigenvalues of these Hamiltonians are
\be\label{ev2}
\Lambda_R^{(k)}=\Schur_{[1^k]}({\bf p}_m^{(R)})
\ee
where
\be\label{tv}
{\bf p}_m^{(R)}:=\{t^m/q^m\}+\{t^m\}\{t^m/q^m\}\sum_i t^{(1-2i)m}\Big(q^{2mR_i}-1\Big)=
\{t^m\}\{t^m/q^m\}\sum_i t^{(1-2i)m}q^{2mR_i}
\ee
The generating function of these eigenvalues can be written in the form
\be
\sum_k\Lambda_R^{(k)}z^k=\prod_{i=1}^{l_{_R}}{\Big(1+zq^{2R_i-1}t^{3-2i}\Big)\Big(1+zq^{2R_i+1}t^{-1-2i}\Big)
\over \Big(1+zq^{2R_i-1}t^{1-2i}\Big)\Big(1+zq^{2R_i+1}t^{1-2i}\Big)}
\ee
The Hamiltonians (\ref{10}) are related with these by the triangle transformation
\be
\hat {\cal H}_1=\{t/q\}\hat H_1,\nn\\
\hat {\cal H}_2={\{q^2\}\{q/t\}\over\{q\}\{t\}}\hat H_1^2+{1\over 2}\{q^2/t^2\}\hat H_2\nn\\
\hat {\cal H}_3={\{q^3\}\{t/q\}\over\{q\}\{t\}^2}\hat H_1^3+{1\over 2}{\{t/q\}\over\{t^2\}}
\Big({\{q^2\}\{t^2\}\{q^2/t^2\}\over\{q\}\{t\}\{q/t\}}+{\{q\}\{q^6\}\over\{q^2\}\{q^3\}}\Big) \hat H_1\hat H_2-{1\over 3}\{q^3/t^3\}\hat H_3\nn\\
\ldots
\ee

\bigskip

One can also make another triangle transformation in order to construct new Hamiltonians $\mathfrak{H}_k$, their generation function being
\be
\sum_k{(-1)^{k+1}\over k}\hat{\mathfrak{H}}_kz^k:=\log\Big(\sum_k\hat{\cal H}_kz^k\Big)
\ee
The eigenvalues of these new Hamiltonians are $\mathfrak{L}_R^{(k)}={\bf p}_k^{(R)}$. This means that they celebrate the property
\be\label{efM}
\boxed{
\Mac_P(\hat{\mathfrak{H}}_k)\cdot \Mac_R(p)=\Mac_P({\bf p}_k^{(R)})\cdot \Mac_R(p)}
\ee

Note that one can choose various generating functions for the Hamiltonians (\ref{Ham2}), e.g. that of the exponential type
\be
F_H(z):={1\over g_1}\sum_{k=0}{\hat{\cal H}_{k+2}\over k!}\Big(z\{q\}\{t\}\Big)^k=[e^{-z\hat P_0}\hat P_{-1}e^{z\hat P_0},\hat P_1]
\ee
This generating function does not include $\hat {\cal H}_1\sim \hat P_0$.

\subsection{Macdonald polynomials: summary}

Thus, we explained that the Macdonald polynomials can be unambiguously defined either by the triangle expansion (\ref{PthroughSchurs}) with the requirement of the orthogonality condition w.r.t. the scalar product (\ref{sp}), or by the requirement that they are eigenfunctions of the Hamiltonian (\ref{Ham1}), which comes from the Ruijsenaars-Schneider integrable system.
However,  {\bf triangularity is not yet apparent from the shape of the Hamiltonian}
and this relation requires better understanding,
see \cite{triang} for a possible approach to this problem.

Because of integrability, in this system there are also higher Hamiltonians.
No canonical way is still available to construct them from the first principles.
We mention just three possible approaches.

One of the possibilities is the Hamiltonians (\ref{10}), which have their eigenvalues (\ref{ev1}) expressed through linear combinations of the antisymmetric Schur (or, which is the same, of antisymmetric Macdonald polynomials), the latter being symmetric polynomials of the variables (\ref{sv1}). Let us note that the shifted (or interpolating) antisymmetric Macdonald polynomials \cite{shifted} are linear combinations very similar to (\ref{ev1}):
\be
\Mac^*_{[1^n]}\{\tilde p_k\}=\sum_{k=0}^{n-1}
{\prod_{j=1}^m\Big(t^{2(k-j+1)}-1\Big)\over \prod_{j=1}^m\Big(t^{2j}-1\Big)}\cdot \Mac_{[1^{k+1}]}\{\tilde p_k\}
\ee
where
\be
\tilde p_k:=\sum_i (x_i^k-1)t^{-2ik}
\ee
The difference is only in an additional product in the denominator of the summand.

Another way of choosing the Hamiltonians is (\ref{Ham2}), they have the eigenvalues (\ref{ev2}), which are just antisymmetric Schur or Macdonald polynomials (without the shift). These polynomials are functions of four sets of time-variables
\be
p_{k,a}^{R}=(-1)^a\beta_{a}^k\sum_iq^{2kR_i}t^{(1-2i)k}
\ee
with four different coefficients $\beta_{1,3}=q^{\pm 1}$, $\beta_{2,4}=(t^2/q)^{\pm 1}$, (\ref{tv}). Such time-variables typically emerge in issues related to the refined topological vertex \cite[formula (52)]{AKMM}. Moreover, the structure of expression like the r.h.s. of (\ref{efM}) is much similar to that of the Hopf hyperpolynomial, \cite[formula (39)]{AKMM}, though that expression depends only on two sets of time variables, \cite[formula (32)]{AKMM}
\be
{\bf p}^{*\bar R}_k=\pm A^{\pm k}\sum_iq^{2kR_i}t^{(1-2i)k}
\ee
Here $\bar R$ denotes the representation of $SL_N$ conjugate to $R$.

The third approach is mentioned in \cite{triang}:
the first Hamiltonian converts Schur functions into bilinear sums
\be
\chi_R \longrightarrow \sum_{h_X=h_Y=1} \alpha_{_X}\beta_{_Y}\chi_{_X} \chi_{_{R/Y}}
\ee
restricted to {\it single}-hook Young diagrams $X$ and $Y$ of same size.
In higher Hamiltonians restriction is weakened to $k$-hook diagrams.
This approach has chances of revealing  the origins of triangularity and could
allow generalization from the Schur polynomials not only to the Macdonald ones, but also to the Kerov functions,
but there is long way to work all this out.

\section{Introduction to GMP}

Let us now discuss if one can define the GMP using one of the two possibilities that we discussed in the previous section for the ordinary Macdonald polynomials.

\subsection{Triangular structure of GMP}

As reviewed in detail in recent \cite{MMkerov},
the basic property of all symmetric polynomials naturally emerging in the course of study of non-perturbative quantum field theory,
is that they are triangular combinations of Schur functions.
This property is a bonus from existence of the natural lexicographical
ordering for Young diagrams, and
actually it breaks down when one goes up from ordinary to plane partitions
\cite{3Schurs}, but it is still true for finite sets
of Young diagrams, and thus for the Generalized Macdonald and Kerov functions.

\paragraph{Triangular expansion.} The GMP $\Mac_{R_1,\ldots,R_n}\{p_k^{(1)}, \ldots, p_k^{(n)}|A_1,\ldots,A_n\}$
depends on an ordered sequence of $n$ Young diagrams, on $n$
infinite sets of time variables and on $n$ ``deformation" parameters $A_i$,
sometimes additionally constrained by the condition $\prod_{i=1}^n A_i = 1$:
Since nothing especially interesting depends on $n$,
we mostly consider the case of $n=2$ in order to simplify formulas, with a single parameter $A$ (usually substituted by $Q=A^2$)
such that $A_1=A$, $A_2=A^{-1}$, and denote $p^{(1)}_k=p_k$, $p^{(2)}_k=\bar p_k$,
i.e. our typical notation will be $\Mac^{(Q)}_{R_1,R_2}\{p,\bar p\}$. Similarly to the ordinary Macdonald polynomial case, we present the GMP as a triangular linear combination of the products of the corresponding Schur functions:
\be\label{trGMP}
\Mac_{R_1,R_2}\{p,\bar p\} 
=\Schur_{R_1}\{p\}\Schur_{R_2}\{\bar p\}
+ \sum_{(R'_1,R'_2)<(R_1,R_2)} {\CC}_{R_1,R_2|R'_1,R'_2}(A,q,t)\cdot
\Schur_{R'_1}\{p\}\Schur_{R'_2}\{\bar p\}
\ee
Let us discuss the ordering in this sum.
At each level $L=|R_1|+|R_2|$, the pairs of Young diagrams are ordered in such a way that
$R_2$ is ordered first. The ordering starts from diagrams of smaller sizes and is lexicographical  for the same size diagrams. Then, one similarly orders $R_1$ within diagrams with coinciding $R_2$.
For instance, at level $2$:
\be
[1,1],\emptyset; \ \ \ [2],\emptyset; \ \ \ [1],[1]; \ \ \ \
\emptyset,[1,1]; \ \ \ \emptyset,[2]
\ee
In Kerov generalizations \cite{Kerov}, a role is played by a transposed lexicographical ordering (see \cite{MMkerov} for details),
but, in the Macdonald case, the difference between these orderings is inessential as we explained in the previous section,
and this remains true for the GMP. Moreover, it is sufficient to consider the partial ordering
\be
(R^{(1)},R^{(2)})>(P^{(1)},P^{(2)})\ \ \ \hbox{if}\ \ \ \ \left\{
\begin{array}{cr}\sum_{i=1}^k(R^{(2)}_i-P^{(2)}_i)>0&\forall k
\\ &
\\ |R^{(2)}|-|P^{(2)}|+\sum_{i=1}^k(R^{(1)}_i-P^{(1)}_i)>0&\forall k
\end{array}\right.
\ee
and, in all cases when the diagrams can not be ordered, the corresponding Kostka-Macdonald coefficients ${\CC}_{R_1,R_2|R'_1,R'_2}(A,q,t)$ in (\ref{trGMP}) vanish.

Note that, from the triangular structure of the usual Macdonald polynomials, it follows that there is also an expansion
\be
\Mac_{R_1,R_2}\{p,\bar p\} =
\Mac_{R_1}\{p\}\Mac_{R_2}\{\bar p\}
+ \sum_{(R'_1,R'_2)<(R_1,R_2)} {\C}_{R_1,R_2|R'_1,R'_2}(A,q,t)\cdot
\Mac_{R'_1}\{p\}\Mac_{R'_2}\{\bar p\}
\label{triexpan}
\ee
For the purposes of the present paper the most convenient is this expansion in the Macdonald polynomials, not in the Schur ones.

The GMP are defined so that at the beginning of the lexicographic sequence,
when $R_2=\emptyset$,
they do not depend on $Q=A^2$ and coincide with the ordinary Macdonald polynomials,
\be
\Mac^{(Q)}_{R,\emptyset}\{p,\bar p\} = \Mac_R\{p\}
\ee
(note that the ``complementary" $\Mac^{(Q)}_{\emptyset,R}\{p,\bar p\}$ is a full-fledged
function of both time variables and $A$, and has not much to do with $\Mac_{R}\{\bar p\}$).
Another reduction appears in the un-deformed case, for $A=\infty$:
\be
\Mac^{(\infty)}_{R_1,R_2}\{p,\bar p\} = \Mac_{R_1}\{p\}\cdot\Mac_{R_2}\{\bar p\}
\label{Minf}
\ee

\paragraph{The scalar product.}

Now one has to specify the transform, i.e. to explain how the coefficients
$C$ are defined.
In \cite{MMkerov} for the ordinary Macdonald and Kerov functions, we imposed an
orthogonality condition, but for the GMP this is not immediate:
one should know what the relevant scalar product is.
Our goal is to construct such a scalar product,
and thus to put the theory of GMP into the general context of \cite{MMkerov},
but this requires some work and insights. Before doing this, let us note that there is a standard scalar product for the GMP, however, the system of GMP's is not orthogonal, but bi-orthogonal w.r.t. it: one has to introduce a dual set of the GMP

The standard product is an independent product of those for $p_k$ and $\bar p_k$, (\ref{sp}):
\be
\Big< { p}_{\Delta_1}\bar{ p}_{\Delta_2}\Big| { p}_{\Delta'_1}\bar{ p}_{\Delta_2'} \Big>_{(q,t)}=
z_{\Delta_1}z_{\Delta_2} \cdot \delta_{\Delta_1,\Delta_1'}\delta_{\Delta_2,\Delta_2'}\cdot
\left(\prod_{i=1}^{l_{\Delta_1}} {\{q^{{\delta_1}_i}\}\over\{t^{{\delta_1}_i}\}}\right)
\left(\prod_{i=1}^{l_{\Delta_2}} {\{q^{{\delta_2}_i}\}\over\{t^{{\delta_2}_i}\}}\right)
\ee
The ordinary Macdonald functions are orthogonal in this scalar product,
\be
\Big< \Mac_{R_1}\{p\}\Mac_{R_2}\{\bar p\}\Big|\Mac_{R'_1}\{p\}  \Mac_{R'_2}\{\bar p\}\Big> =
||\Mac_{R_1}||^2||\Mac_{R_2}||^2\cdot \delta_{R_1,R'_1}\cdot \delta_{R_2,R_2'}
\label{scapr}
\ee
but  generalized Macdonald polynomials $\GMac_{R_1,R_2}\{p,\bar p\}$
are orthogonal to {\it another} set of ``dual" operators
$\widetilde \GMac_{R_1,R_2}\{p,\bar p\}\neq \GMac_{R_1,R_2}\{p,\bar p\}$,
obtained from $\Mac_{R_1}\{p\}\Mac_{R_2}\{\bar p\}$ by the transposed triangular transform:
\be
\Big< \widetilde \GMac_{R_1,R_2}\{p,\bar p\}\Big|\GMac_{R'_1,R'_2}\{p,\bar p\}\Big> =
||\Mac_{R_1}||^2||\Mac_{R_2}||^2\cdot \delta_{R_1,R'_!}\cdot \delta_{R_2,R'_2}
\label{orthodualGMP}
\ee
The problem is that the freedom
is not fixed by {\it such} orthogonality requirement, and there is
no guiding principle to select the needed triangular transformation.

Let us note that the transformation from the Schur to Macdonald polynomials
is lower triangular both in the normal and dual sectors,
while the further transformation to the GMP continues to be lower
triangular in normal sector, but becomes upper triangular in the dual one.
Thus the net transform from the Schur polynomials to the dual GMP's is not really triangular.

\paragraph{Examples of the triangular expansions.}

Let us look at the first examples:
\be
\GMac_{[1],\emptyset}\{p,\bar p\} = \Mac_{[1]}\{p\}
& & \GMac_{\emptyset,[1]}\{p,\bar p\} = \Mac_{[1]}\{\bar p\}
- \frac{t}{ q}\frac{\{q/t\}}{Q-1}\cdot \Mac_{[1]}\{p\}
\nn \\
\widetilde \GMac_{[1],\emptyset}\{p,\bar p\} =
\Mac_{[1]}\{ p\}
+ \frac{t}{ q}\frac{\{q/t\}}{Q-1}\cdot \Mac_{[1]}\{\bar p\} &
& \widetilde \GMac_{\emptyset,[1]} =  \Mac_{[1]}\{\bar p\}
\ee

\be
\GMac_{[1,1],\emptyset}\{p,\bar p\} = \Mac_{[1,1]}\{p\}\nn \\
\GMac_{ [2],\emptyset}\{p,\bar p\} = \Mac_{[2]}\{  p\} \nn \\
\GMac_{[1],[1]}\{p,\bar p\} = \Mac_{[1]}\{p\}\Mac_{[1]}\{\bar p\}
- \frac{t}{q}\frac{ \{q/t\}}{Qq^{2}-1}\cdot\Mac_{[2]}\{  p\}
-\frac{t}{q}\frac{\{q\}\{t^2\}\{q/t\}}{\{t\}\{qt\}(Qt^{-2}-1)} \Mac_{[1,1]}\{  p\}
\nn \\
\!\!\!\!\!\!\!\!\!\!\!
  \GMac_{ \emptyset,[1,1] }\{p,\bar p\} = \Mac_{[1,1]}\{\bar p\}
-\frac{t}{q}\frac{\{q/t\}}{Qt^2-1}\Mac_{[1]}\{p\}\Mac_{[1]}\{\bar p\}
+ \frac{t^3}{q^3}\frac{\{q/t\}}{Qt^2-1}\Mac_{[2]}\{  p\}
- \nn \\
-\frac{t}{q}\frac{\{q/t\}}{\{t\}\{qt\}}\left(
\frac{t^3}{q}\frac{\{q\}\{q/t\}}{Qt^2-1}
- \frac{t}{q}\frac{\{qt\}\{q/t^2\}}{Q-1}
\right)\Mac_{[1,1]}\{  p\}
\nn\\
 \GMac_{\emptyset, [2] }\{p,\bar p\} = \Mac_{[2]}\{\bar p\}
-\frac{t}{q}
\frac{\{t\}\{q^2\}\{q/t\}}{\{q\}\{qt\}(Qq^{-2}-1)}  \Mac_{[1]}\{p\}\Mac_{[1]}\{\bar p\}
-\nn \\
-\frac{t}{q}\frac{\{q/t\}}{\{q\}\{qt\}}\left(
-\frac{t}{q^3}\frac{\{t\}\{q/t\}}{Qq^{-2}-1} + \frac{t}{q}\frac{\{qt\}\{q^2/t\}}{Q-1}
\right)\Mac_{[2]}\{  p\}
+ \frac{t^3}{q^3}\frac{\{q/t\} \{t^2\}\{q^2\}}{ \{qt\}^2(Qq^{-2}-1)}\Mac_{[1,1]}\{  p\}
\label{lev2GEMs}
\ee
 
\be
\widetilde \GMac_{\emptyset,[2]}\{p,\bar p\} = \Mac_{[2]}\{\bar p\} \nn \\
\widetilde \GMac_{\emptyset,[1,1]}\{p,\bar p\} = \Mac_{[1,1]}\{\bar p\}\nn \\
\widetilde \GMac_{[1],[1]}\{p,\bar p\} = \Mac_{[1]}\{p\}\Mac_{[1]}\{\bar \}
+ \frac{t}{q}\frac{ \{q/t\}}{Qq^{-2}-1}\cdot\Mac_{[2]}\{\bar p\}
+\frac{t}{q}\frac{\{q\}\{t^2\}\{q/t\}}{\{t\}\{qt\}(Qt^2-1)} \Mac_{[1,1]}\{\bar p\}
\nn \\
\!\!\!\!\!\!\!\!\!\!\!\!\!\!\!\!\!\!\!\!\!\!\!\!\!\!\!\!\!\!\!\!\!\!\!\!\!\!\!\!\!\!\!\!\!\!\!\!
\widetilde \GMac_{ [2],\emptyset}\{p,\bar p\} = \Mac_{[2]}\{p\}
+ \frac{t}{q}\frac{\{t\}\{q^2\}\{q/t\}}{\{q\}\{qt\}(Qq^2-1)}\Mac_{[1]}\{p\}\Mac_{[1]}\{\bar p\}
+ \nn \\
+ \frac{\{q/t\}}{\{q\}\{qt\}}
\left(-\frac{t^2\{t\}\{q/t\}}{Qq^2-1} + \frac{t^2}{q^2}\frac{\{qt\}\{q^2/t\}}{Q-1}
\right)\Mac_{[2]}\{\bar p\}
-\frac{t}{q}\frac{\{q^2\}\{t^2\}\{q/t\}}{\{qt\}^2(Qq^2-1)}  \Mac_{[1,1]}\{\bar p\}
\nn \\
\widetilde \GMac_{ [1,1],\emptyset}\{p,\bar p\} = \Mac_{[1,1]}\{p\}
+\frac{t}{q}\frac{\{q/t\}}{Qt^{-2}-1} \Mac_{[1]}\{p\}\Mac_{[1]}\{\bar p\}
-\frac{t}{q}\frac{\{q/t\}}{Qt^{-2}-1} \Mac_{[2]}\{\bar p\}
+ \nn \\
+  \frac{\{q/t\}}{\{t\}\{qt\}}\left(
\frac{1}{q^2}\frac{\{q\}\{q/t\}}{Qt^{-2}-1} - \frac{t^2}{q^2}\frac{\{qt\}\{q/t^2\}}{Q-1}
\right)\Mac_{[1,1]}\{\bar p\}
\ee

One can see in this example that
the GMP themselves can be obtained by a triangular transformation from the Schur bilinears,
\be
\GMac_{[1,1],\emptyset}=\Schur_{[1,1],\emptyset}\nn \\
\GMac_{[2],\emptyset}=\Schur_{[2],\emptyset}
- \frac{ \Big\{\frac{q}{t}\Big\}}{\{qt\}}\cdot \Schur_{[1,1],\emptyset} \nn \\
\GMac_{[1],[1]} = \Schur_{[1],[1]}
- \frac{\frac{t}{q}\Big\{\frac{q}{t}\Big\}}{Qq^2-1}\cdot \Schur_{[2],\emptyset}
- \frac{\frac{t}{q}\Big\{\frac{q}{t}\Big\}
\Big(Qq^2t^2+Qq^2-Qt^2-t^2\Big)}{(Q-t^2)(Qq^2-1)}\cdot \Schur_{[1,1],\emptyset}\nn \\
\GMac_{\emptyset,[1,1]}=\Schur_{\emptyset,[1,1]}
- \frac{\frac{t}{q}\Big\{\frac{q}{t}\Big\}}{Qt^2-1}\cdot \Schur_{[1],[1]}
- \frac{\frac{t^3}{q^3}\Big\{\frac{q}{t}\Big\}}{Qt^2-1}\cdot \Schur_{[2],\emptyset}
- \frac{\frac{t}{q}\Big\{\frac{q}{t}\Big\}(Qt^4-q^2)}{q^2(Qt^2-1)(Q-1)}
\cdot \Schur_{[1,1],\emptyset} \nn \\
\GMac_{\emptyset,[2]}=\Schur_{\emptyset,[2]} - \frac{ \{q/t\}}{\{qt\}}\cdot \Schur_{\emptyset,[1,1]}
- \frac{\frac{t}{q}\Big\{\frac{q}{t}\Big\}\Big(Qq^2t^2-Qq^2+Qt^2-q^2\Big)}
{(Qt^2-1)(Q-q^2)}\cdot\Schur_{[1],[1]} -\nn \\
- \frac{\frac{t}{q}\Big\{\frac{q}{t}\Big\}\Big(Q^2q^2t^2-Qq^4t^2-Qq^2t^2+Qt^4-Qt^2+q^4\Big)}
{q^2(Qt^2-1)(Q-1)(Q-q^2)}\cdot\Schur_{[2],\emptyset}
+ \nn \\
+ \frac{\frac{t}{q}\Big\{\frac{q}{t}\Big\}\Big(Q^2q^2t^4-Qq^2t^4-Qq^4 +Qq^2t^2-Qt^4+q^2t^2\Big)}
{q^2(Qt^2-1)(Q-1)(Q-q^2)}\cdot\Schur_{[1,1],\emptyset}
\label{lev2GEMsviaSchur}
\ee
but it is not the case for the dual GMP.
This is obvious already for the very first one:
in variance with $\GMac{[1,1],\emptyset}\{p,\bar p\} = \Schur_{[1,1]}\{p\}$,
the dual $\widetilde\GMac_{\emptyset,[2]}$
does {\it not} coincide with the corresponding Schur $\Mac_{[2]}\{\bar p\}$.
As already mentioned, the reason for this difference is that the GMP's are obtained
by composing two upper triangular transformation,
but the dual GMP's are compositions of lower and upper triangular transformations,
which are no  longer triangular themselves.

\subsection{GMP as eigenfunctions of deformed Hamiltonians}

Since the GMP are not orthogonal in the standard metric, the Hamiltonians that have the GMP's their eigenfunctions
are not Hermitian in the standard metric, though the
deformation is triangular.
The typical form is $p\frac{\p}{\p \bar p}$,
so that the ordinary Macdonald polynomials $\GMac_{R,\emptyset}\{p,\bar p\} = \Mac_R\{p\}$
are not affected.

In fact, it is important to add some additional Hermitian pieces,
and they bring into the game the deformation parameter(s).
For example,
\be
up\p_p + (\bar u\bar p +\epsilon p) \p_{\bar p}
\ee
has eigenfunctions $p$ and $\bar p - \frac{\eta}{u-\bar u}\cdot p$
with eigenvalues $u$ and $\bar u$.
It becomes Hermitian if the scalar product is deformed to
\be
\Big<p\Big|\bar p\Big> \ =\  -\frac{\eta}{u-\bar u}
\label{defoscasample}
\ee
In general, $u$ is a coefficient in front of $np_n\frac{\p}{\p p_n}$
(or higher Macdonald Hamiltonians), and this term just shifts
the eigenvalue of $\Mac_R\{p\}$, while the mixing term
$\sim \epsilon_n\cdot n^2p_n\frac{\p}{\p\bar p_n}$
annihilates it.
However, the $\bar p$-dependent eigenfunctions are no longer just
$\Mac_{R_1}\{p\}\Mac_{R_2}\{\bar p\}$ with $R_2\neq \emptyset$, they
get non-trivially $\epsilon$-deformed into the GMP.

In fact, it is sufficient to consider only one Hamiltonian, it fully
determines the deformation.
All the higher Hamiltonians are deformed in a consistent way.

\paragraph{Perturbation theory.}
The main point is that perturbation theory is greatly simplified for triangular perturbations.
If the deformed Hamiltonian is $\hat H+\hat V$ and $\hat H\psi_i = \lambda_i\psi_i$,
then
\be
(\hat H+\hat V)\left(\psi_i+\sum_{k<i} \alpha_i^k\psi_k\right) =
\lambda_i \psi_i + \sum_{k<i} \alpha_i^k \lambda_k \psi_k + \underline{\sum_{k<i}}
V^k_i \psi_k + \underline{\sum_{l<k<i}}    \alpha_i^l V_l^k \psi_k
= \lambda_i\left(\psi_i+\sum_{k<i} \alpha_i^k\psi_k\right)
\ee
and
\be
\alpha_{i}^k = \frac{1}{\lambda_i-\lambda_k}\left(V_i^k + \sum_{l<k}\alpha_i^l V _l^k\right)
= \frac{V_i^k}{\lambda_i-\lambda_k}+ \sum_{l<k}
\frac{V_i^lV_l^k}{(\lambda_i-\lambda_k)(\lambda_i-\lambda_l)}
+ \sum_{m<l<k}
\frac{V_i^mV_m^lV_l^k}{(\lambda_i-\lambda_k)(\lambda_i-\lambda_l)(\lambda_i-\lambda_m)}
+ \ldots
\nn
\ee
The underlined sums are triangular, if such is the perturbation matrix: $V_i^k\neq 0$
only for $k<i$.
In this case, the eigenvalues do not change, and the eigenfunctions are finite sums with
$V$ entering the expression for the $i$-th eigenfunctions at most in the $i$-th power.

In application to the GMP, the role of $\psi_i$ is played by $\Mac_R\bar \Mac_Q$,
and the operator $\hat V$ written symbolically as $\hat V = \sum_{k}V_{k}p_k\frac{\p}{\p\bar p_k}$ decreases the size of $Q$ and enlarges that of $R$:
\be
\hat V \, (\Mac_R\bar \Mac_Q) = \sum_{m=1} \sum_{\stackrel{R'\vdash |R|+m}{Q'\vdash |Q|-m}}
V_{R,Q}^{R',Q'} \Mac_{R'}\bar \Mac_{Q'}
\ee
which brings $(R,Q)$ down in the ordering.
The new eigenfunctions, i.e. exactly the GMP, are given by
\be
\GMac_{R,Q} =
\Mac_R\bar \Mac_Q + \CK_{R,Q}^{R',Q'} \Mac_{R'}\bar \Mac_{Q'}
\ee
with
\be
\CK_{R,Q}^{R',Q'} = \frac{V_{R,Q}^{R',Q'}}
{\lambda_{R,Q} - \lambda_{R',Q'}}
 + \sum_{(R',Q')<(R'',Q'')<(R,Q)}
\frac{V_{R,Q}^{R'',Q''}V_{R'',Q''}^{R',Q'}}
{(\lambda_{R,Q} - \lambda_{R',Q'})(\lambda_{R'',Q''} - \lambda_{R',Q'})} + \ldots
\label{Kmatrels}
\ee
The eigenvalues actually have the form
\be
\lambda_{(R_1,R_2)} = \lambda_{R_1} + Q^{-1} \lambda_{R_2}
\ee
with
\be
\lambda_R =  \sum_{i=1}^{l_R} \frac{q^{2r_i}-1}{t^{2i}}
\ee

Perturbed eigenfunctions are no longer orthogonal in the original scalar
product, where $\Big<\psi_i\Big|\psi_j\Big> = \delta_{i,j}$.
However, one can introduce a new one, where they are, and triangularity
allows one to reformulate it as a simple recursion relation:
\be
\Big<\Big< \psi_i + \sum_{k<i} \alpha_i^k \psi_k \Big|
\psi_j + \sum_{l<j} \alpha_j^l \psi_l\Big>\Big> = \delta_{ij} \ \ \ \ \ \ \
\Longleftrightarrow \nn \\ \Longleftrightarrow \ \ \ \ \ \
\Big<\Big< \psi_i \Big| \psi_j\Big>\Big> =
-\sum_{k<i} \alpha_i^k   \Big<\Big<\psi_k \Big|\psi_j\Big>\Big>
-\sum_{l<j} \alpha_j^l   \Big<\Big<\psi_i \Big|\psi_l\Big>\Big>
-\sum_{k<i}\sum_{l<j}   \alpha_i^k \alpha_j^l  \Big<\Big<\psi_k \Big|\psi_l\Big>\Big>\ \ \ \ \ \ \hbox{for}\ i\ne j
\ee
In particular, for the lowest eigenfunction $\psi_0$, which remains un-deformed,
\be
\Big<\Big< \psi_i + \sum_{k<i} \alpha_i^k \psi_k \Big|\psi_0\Big>\Big> =
\delta_{i0} - \frac{V_i^0}{\lambda_i-\lambda_0} +
\sum_{k<i}\frac{V_i^kV_k^0}{(\lambda_i-\lambda_0)(\lambda_k-\lambda_0)}
+ \sum_{l<k<i}\frac{V_i^kV_k^lV_l^0}
{(\lambda_i-\lambda_0)(\lambda_k-\lambda_0)(\lambda_l-\lambda_0)}
- \ldots
\ee
This expansion is very similar to above one for the eigenfunction itself,
the only differences are alternated signs and the ``common eigenvalue" in denominators:
it is $\lambda_i$ for the eigenfunctions and $\lambda_0$ for the
deformed scalar product.
In the case of GMP, the role of $\psi_0$ is played by {\it any} of the ordinary
$\bar p$-independent Macdonald functions $M_R\{p\}$.

\paragraph{GMP Hamiltonian.}
The actual Hamiltonian in the case of GMP is
\be
\hat {\cal H} (F)\{p,\bar p\} = \frac{1}{t^2-1}\left\{-(1+Q^{-1})\cdot F\{p,\bar p\} +
{\rm res}_{z=0}\left(
\exp\left(\sum_n \frac{(1-t^{-2n})p_nz^n}{n}\right)
F\left\{p_k+\frac{q^{2k}-1}{z^k},\bar p_k\right\}
+ \right.\right. \nn \\ \!\!\!\!\!\!\!\!\!\!\!\!\! \left.\left.
+ Q^{-1}\cdot \exp\left(\sum_n \frac{(1-t^{-2n})z^n}{n}
\Big(\epsilon_np_n+\bar p_n)\right)
F\left\{p_k,\bar p_k + \frac{q^{2k}-1}{z^k} \right\}
\right)\right\} \ \ \ \ \
\label{GEMham}
\ee
and perturbation is generated by the $\epsilon_n$-term, $\epsilon_n:=1-(t/q)^{2n}$.
Thus it looks more like
\be
\hat{\cal H} = \hat H_{\Mac}+\frac{1}{Q}\widehat{\bar H}_{\Mac}
+ \frac{1}{Q}e^{\hat V} \widehat{\bar H}_{\Mac}
\ee
Explicit shifts of the arguments can be substituted by action of the operators like
\be
F\left\{p_k+\frac{q^{2k}-1}{z^k},\bar p_k\right\} =
\exp\left(\sum_n \frac{q^{2n}-1}{z^n}\frac{\p}{\p   p_n}\right)F\{p_k,\bar p_k\}
\ee
The eigenvalue of $\GMac_{R,S}\{p,\bar p\}$ remains un-deformed, i.e. is the same
as it was for $\Mac_R\{p\}\cdot \Mac_S\{\bar p\}$ with the un-deformed Hamiltonian (c.f. with (\ref{1ev})):
\be
\lambda_{R,S} = \sum_{i=1}^{l_R} \frac{q^{2r_i}-1}{t^{2i}} +
Q^{-1} \sum_{j=1}^{l_S} \frac{q^{2s_j}-1}{t^{2j}}
\ee

\paragraph{Higher Hamiltonians.}

Similarly to the case of ordinary Macdonald polynomials, the GMP's are unambiguously defined as graded polynomial eigenfunctions of the first Hamiltonian (\ref{GEMham}). This clearly implies an underlying integrable structure. It simultaneously implies that there exist higher Hamiltonians, and we again borrow formulas from \cite{MMZ} (see also \cite[Appendix B]{AR}). In fact, one can use the form (\ref{Ham2}) for the Macdonald Hamiltonians in order to directly extended them to the GMP case: one just needs to substitute in all formulas $P_m$ with
\be
\hat P^{(2)}_m:=\oint_{z=0}{dz\over z^{m+1}}\hat V^{(2)}(z)
\ee
\be
\hat V^{(2)}(z):=\exp\left(\sum_{k>0}p_k{(1-t^{-2k})\over k}z^k\right)\cdot\exp\left(\sum_{k>0}{q^{2k}-1\over z^k}
{\partial\over\partial p_k}\right)+\nn\\
+{1\over Q}\exp\left(\sum_{k>0}\Big(\bar p_k+\epsilon_kp_k\Big){(1-t^{-2k})\over k}z^k\right)\cdot\exp\left(\sum_{k>0}{q^{2k}-1\over z^k}
{\partial\over\partial \bar p_k}\right)
\ee
with $\epsilon_k:=1-{t^{2k}\over q^{2k}}$.
Then, the Hamiltonians for the GMP are given by the same formulas (\ref{Ham2}), and the eigenvalues are given by formulas similar to (\ref{ev2}), but depending on two Young diagrams:
\be\label{Ham22}
\hat {\cal H}_k^{(2)}={g_1\over\{q\}^k\{t^{-1}\}^k}[[\ldots[\hat P^{(2)}_{-1},\underbrace{\hat P^{(2)}_0],\ldots,
\hat P^{(2)}_0}_{k-2\ times}],\hat P^{(2)}_1]
\ee
\be
\hat {\cal H}_k^{(2)}\cdot \GMac_{(R,P)}=\Lambda_{(R,P)}^{(k)}\cdot \GMac_{(R,P)}
\ee
where
\be\label{evGMP}
\Lambda_{(R,P)}^{(k)}=\sum_{n=0}^kQ^{-n}\Lambda_R^{(k-n)}\Lambda_P^{(n)}=
\sum_{n=0}^k\Schur_{[1^{k-n}]}({\bf p}_m^{(R)})\Schur_{[1^n]}(Q^{-m}{\bf p}_m^{(P)})=
\Schur_{[1^k]}({\bf p}_m^{(R)}+Q^{-m}{\bf p}_m^{(P)})
\ee
is expressed through the eigenvalues (\ref{ev2}).

The generating function for these eigenvalues is given by
\be
\sum_k\Lambda_{(R,P)}^{(k)}z^k=\prod_{i=1}^{l_{_R}}{\Big(1+zq^{2R_i-1}t^{3-2i}\Big)\Big(1+zq^{2R_i+1}t^{-1-2i}\Big)
\over \Big(1+zq^{2R_i-1}t^{1-2i}\Big)\Big(1+zq^{2R_i+1}t^{1-2i}\Big)}\times
\prod_{i=1}^{l_{_R}}{\Big(1+z/Qq^{2P_i-1}t^{3-2i}\Big)\Big(1+z/Qq^{2P_i+1}t^{-1-2i}\Big)
\over \Big(1+z/Qq^{2P_i-1}t^{1-2i}\Big)\Big(1+z/Qq^{2P_i+1}t^{1-2i}\Big)}\nn
\ee
An extension from the bilinear to multilinear GMP is evident.

One can again introduce the Hamiltonians $\hat{\mathfrak{H}}_k^{(2)}$ with the generating function
\be
\sum_k{(-1)^{k+1}\over k}\hat{\mathfrak{H}}_k^{(2)}z^k:=\log\Big(\sum_k\hat{\cal H}_k^{(2)}z^k\Big)
\ee
with the eigenvalues $\mathfrak{L}_{(R,P)}^{(k)}={\bf p}_m^{(R)}+Q^{-m}{\bf p}_m^{(P)}$, and
\be\boxed{
\Mac_S(\hat{\mathfrak{H}}_k^{(2)})\cdot \GMac_{(R,P)}(p)=\Mac_S({\bf p}_m^{(R)}+Q^{-m}{\bf p}_m^{(P)})\cdot
\GMac_{(R,P)}(p)}
\ee

One more notable thing is emergency of the {\it triple} of parameters
$(q_1,q_2,q_3) = (q,t^{-1},tq^{-1})$, which is not seen at the level of
the ordinary Macdonald functions. Thus the GMP are better suited for studying
the apparent triality of the DIM algebra \cite{triality}.

\subsection{GMP: summary}

Thus, we studied the Macdonald polynomial features that could be used for their definition and that can be extended to the GMP case. In particular, there is still a triangular structure of expansion, (\ref{trGMP}) in the Schur polynomials. Moreover, in this case, there is also a triangular expansion in Macdonald polynomials (\ref{triexpan}). However, there is no known scalar product so far that could fix the triangular expansion coefficients from the orthogonality condition.

There is a Hamiltonian (\ref{GEMham}) whose eigenfunctions are the GMP's, and it unambiguously defines them. However, this Hamiltonian at the moment comes only from an additional algebraic set-up of quantum toroidal algebras \cite{Awata,Ohkubo}. Moreover, in a complete analogy with the standard Macdonald case, one can construct higher Hamiltonians (\ref{Ham22}) with eigenvalues given by the same antisymmetric Schur or Macdonald polynomials. However, they are now polynomials of a sum of two sets of time-variables (\ref{evGMP}), each of them being associated its own Young diagram in accordance with the same Macdonald case formula (\ref{tv}).

Note that such a sum of times corresponding to two different Young diagrams emerges within the same context of refined topological vertex and Hopf hyperpolynomial as we discussed in s.2.3, however, in the case of composite representations (see \cite{MMc,AKMM}, in particular, formula (32) in \cite{AKMM}).

In the next sections, we are going to modify
the Macdonald scalar product so that the $Q$-deformed Hamiltonians become Hermitian.
With this scalar product, the generalized Macdonald functions are themselves orthogonal,
there is no need for the dual set of functions.
If one could define (select) this deformed product by some
appealing and simply formulated {\it ansatz},
this would fix the triangular transform and thus define
the generalized Macdonald polynomials without any direct reference
to the deformed Hamiltonians.
We will look for this scalar product basing on the known set
of GMP's, and then reverse the logic:
we {\it postulate} it as a new {\it definition} of the GMP.

Our other goal in the forthcoming sections is to clarify reasons why the GMP Hamiltonian (\ref{GEMham}) is distinguished without references to the algebraic set-up. On this way, we will construct another set of symmetric functions of two sets of variables, which we call generalized Schur functions.

\section{Looking for a scalar product}

Let us discuss what could be the scalar product that gives rise to the correct GMP from the orthogonality relations. Let us denote
\be
\kappa:=1-\frac{t^2}{q^2} = \frac{t}{q}\Big\{\frac{q}{t}\Big\}\nn\\
G_{R_1,R_2|S_1,S_2} = \Big<\Big< \Mac_{R_1}\{p\}\Mac_{R_2}\{\bar p\}\Big|
\Mac_{S_1}\{p\}\Mac_{S_2}\{\bar p\}\Big>\Big>\nn\\
{\cal G}_{R_1,R_2} = G_{R_1,R_2|R_1,R_2}
\ee
Now we require the orthogonality of the GMP,
\be
\Big<\Big< \GMac_{(R_1,R_2)}\{p,\bar p\}\Big|
\GMac_{(S_1,S_2)}\{p,\bar p\}\Big>\Big>\sim\delta_{R_1,S_1}\delta_{R_2,S_2}
\ee
In particular,
\be
\Big<\Big< \GMac_{(R,\emptyset)}\{p,\bar p\}\Big|
\GMac_{(S,\emptyset)}\{p,\bar p\}\Big>\Big> = \Big<\Big< M_{R}\{p\}|M_S\{p\}\Big>\Big>=G_{R,\emptyset|R,\emptyset}
= {\cal G}_{R,\emptyset}\cdot \delta_{R,S}=||\Mac_R||^2\cdot \delta_{R,S}
\ee
for all $R$ and $S$ at all levels.

Now let us proceed at the first two levels. At the first level, there is only one non-diagonal item of the symmetric matrix $G_{R_1,R_2|S_1,S_2}$:
\be
G_{\emptyset,[1]\big|[1],\emptyset} = \frac{\kappa\cdot {\cal G}_{[1],\emptyset}}{Q-1}
&\ \ \ \ \ {\rm while} & ||M_{[1]}||^2 = \frac{\{q\}}{\{t\}}
\ee
This defines the scalar product completely.

At level $2$, there are more relations, but many more non-diagonal matrix elements of $G_{R_1,R_2|S_1,S_2}$:
\be
G_{[1],[1]\big|[1,1],\emptyset} = \frac{\{q\}\{t^2\}}{\{t\}\{qt\}}
\cdot\frac{\kappa\cdot {\cal G}_{[1,1],\emptyset}}{Qt^{-2}-1}
& \ \ \ \ & ||M_{[1,1]}||^2 = \frac{\{q\}\{qt\}}{\{t\}\{t^2\}}
\nn \\
G_{[1],[1]\big|[2],\emptyset} = \frac{\kappa\cdot {\cal G}_{[2],\emptyset}}{Qq^2-1}
&& ||M_{[2]}||^2 = \frac{\{q\}\{q^2\}}{\{t\}\{qt\}}
\nn \\
G_{\emptyset,[1,1]\big|[1,1],\emptyset} = \frac{\kappa\cdot{\cal G}_{[1,1],\emptyset}}
{\{t\}\{qt\}}\left(\frac{1}{qt}\frac{\{q\}\{q/t\}}{ Qt^{-2}-1}
-\frac{\frac{t}{q}\{qt\}\{q/t^2\}}{Q-1}\right)
\nn \\
G_{\emptyset,[1,1]\big|[2],\emptyset} = \frac{\kappa\cdot {\cal G}_{[2],\emptyset} }{Qq^2-1}&&
\nn \\
G_{\emptyset,[2]\big|[1,1],\emptyset} = \frac{\{q^2\}\{t^2\}}{\{qt\}^2}\cdot
\frac{\kappa\cdot{\cal G}_{[1,1],\emptyset}}{Qt^{-2}-1}
\nn \\
G_{\emptyset,[2]\big|[2],\emptyset} = \frac{\kappa\cdot{\cal G}_{[2],\emptyset}}{\{q\}\{qt\}}
\left(\frac{\frac{t}{q}\{q^2/t\}\{qt\}}{Q-1} - \frac{qt\{t\}\{q/t\}}{Qq^2-1}\right)
\ee
and
\newpage
{\footnotesize
$$
G_{\emptyset,[1,1]\big|[1],[1]} =
\frac{\frac{t^2}{q^2}\cdot\kappa\cdot {\cal G}_{[2],\emptyset}}{Qq^2-1}
- \frac{\frac{t^4}{q^4}\cdot\kappa\cdot {\cal G}_{[2],\emptyset}}{Qt^2-1}
+ \frac{\kappa\cdot {\cal G}_{[1],[1]}}{Qt^2-1}
 + 
 $$
 $$
+ \frac{\{q\}\{t^2\}}{\{t\}^2\{qt\}^2}\cdot
\frac{\kappa^2\cdot {\cal G}_{[1,1],\emptyset}}{\{qt\}^2}
\left( \frac{\frac{t^2}{q}\{qt\}\{q/t^2\}}{\{t\}(Q-1)}
- \frac{\frac{t^5}{q}\{q\}\{q/t\}}{\{t^2\}(Qt^2-1)}
+\underline{\frac{q^2t^8-q^4t^2-q^2t^4+q^2t^2-t^4+q^2}{q^3t^2\{t^2\}(Q-t^2)}}
\right)
$$

\bigskip

$$
G_{\emptyset,[2]\big|[1],[1]} = -\frac{t^4}{q^2}\frac{\{q\}\{q^2\}\{t^2\}^2
\cdot \kappa\cdot {\cal G}_{[1,1],\emptyset}}{\{t\}\{qt\}}
\cdot\left(\frac{1}{Q-q^2}-\frac{1}{Q-t^2}\right)
+ \frac{q^2\{t\}\{q^2\}}{\{q\}\{qt\}}\cdot\frac{\kappa\cdot G_{[1],[1]\big|[1],[1]}}{Q-q^2}
+ 
$$
$$
+ \kappa\cdot{\cal G}_{[2],\emptyset}\left(
-\frac{\kappa\cdot \{t\}}{q^2\{q\}\{q^2\}\{qt\}(Q-q^2)}
+ \frac{t\{q^2/t\}}{q^2\{q\}^2(Q-1)} -
\underline{\frac{q^8t^2+q^6t^2-q^4t^4-q^6-q^4t^2+t^2}{q^4t\{q\}\{q^2\}\{qt\}(Qq^2-1)}}
\right)
$$

\bigskip

$$
G_{\emptyset,[2]\big|\emptyset,[1,1]} = \kappa^2 \left\{
\frac{q\{t\}\{q^2\}{\cal G}_{[1],[1]}}{t\{q\}\{qt\}^2}
\left(\frac{1}{Q-q^2} - \frac{t^2}{Qt^2-1}\right)
+ \right.
$$
$$
\left.
+ \left(-\frac{t\{q^2/t\}}{q^2\{q\}^2(Q-1)}
- \frac{t\{t\}\{q/t\}^2}{q^5\{q\}\{q^2\}\{qt\}^2(Q-q^2)}
-\frac{t^3\{t\}\{q^2\}}{q^3\{q\}\{qt\}^2(Qt^2-1)}
+\underline{\frac{(q^8t^2-q^6t^4+2q^6t^2-q^4t^4-q^6-q^4t^2+q^2t^4-q^2t^2+t^2)}
{q^4t\{q\}\{q^2\}\{qt\}(Qq^2-1)}}
\right){\cal G}_{[2],\emptyset}
+ \right.
$$

$$ 
\left.
+ \left(
-\frac{t^2\{t^2\}\{q/t^2\}}{q\{t\}(Q-1)}
- \frac{t^3\{q\}\{t^2\}^2}{q^3\{qt\}^2(Q-q^2)}
+ \frac{t^7\{q\}\{q/t\}^2}{q\{qt\}^2(Qt^2-1)}
+ \underline{\frac{-q^2t^8+q^2t^6+q^4t^2+q^2t^4-t^6-2q^2t^2+t^4-q^2+t^2}{q^3t^2\{qt\}(Q-t^2)}}
\right)  \frac{\{q^2\}\cdot {\cal G}_{[1,1],\emptyset}}{\{t\}\{qt\}^2}
\right\}
$$
}

\bigskip

The ugly-looking underlined terms can be eliminated or improved
by choosing appropriate deformations of the norms:
\be
{\cal G}_{[1,1],\emptyset} = ||\Mac_{[1,1]}||^2 \cdot \frac{Q-t^2}{Q-q^{-2}},
\ \ \ \ \ \ \ \
{\cal G}_{[2],\emptyset} = ||\Mac_{[2]}||^2\cdot \frac{Q-q^{-2}}{Q-t^2}
\ee
Then
\be
G_{[1],[1]\big|[1,1],\emptyset} = \frac{\kappa}{Qq^2-1}\cdot \frac{q^2t^2\{q\}^2}{\{t^2\}}
\nn \\
G_{[1],[1]\big|[2],\emptyset} = \frac{\kappa}{Q-t^2}\cdot \frac{\{q\}\{q^2\}}{q^2\{t\}\{qt\}}
\nn \\
G_{\emptyset,[1,1]\big|[1,1],\emptyset}
= \frac{\kappa}{t\{t\}\{t^2\}}\left(\frac{t^3\{qt\}\{q/t^2\}}{Q-1}
+ \frac{q^2t^6-q^4t^2+q^2t^4-2t^4+1}{Qq^2-1}\right)
\nn \\
G_{\emptyset,[2]\big|[1,1],\emptyset}
= -\frac{\kappa}{Qq^2-1}\cdot \frac{q^2t^2\{q\}\{q^2\}}{\{t\}\{qt\}}
\nn \\
G_{\emptyset,[2]\big|[2],\emptyset}
=\frac{\kappa\cdot \{q\}\{q^2\}}{q^5t^2\{t\}^2\{qt\}^2}
\left(-\frac{q^3t^2\{qt\}\{q^2/t\}}{Q-1} + \frac{q^6t^4-2q^2t^4-q^4+q^2t^2+t^2}{Q-t^2}\right)
\ee
Now there are two natural versions of the other elements of the scalar product matrix $G_{R_1,R_2|S_1,S_2}$.

\paragraph{Version A:}

Assume that the norms of un-deformed polynomials
$\Mac_R\{p\}$ also remain un-deformed:
\be
{\cal G}_{R,\emptyset} = ||\Mac_R||^2
\ee
Then
\be
\Big<\Big< \bar p_1| p_1 \Big>\Big> = \frac{\kappa}{Q-1} \Big<p_1\Big|p_1\Big>
\nn \\
\Big< p_1\bar p_1| \Mac_{[1,1]}\{p\} \Big>\Big>
= \frac{\kappa}{Qt^{-2}-1} \Big<p_1^2\Big|\Mac_{[1,1]}\Big>
\nn \\
\Big<\Big< p_1\bar p_1| \Mac_{[2]}\{p\} \Big>\Big> = \frac{\kappa}{Qq^2-1}\Big<p_1^2\Big|\Mac_{[2]}\Big>
\ee

\be
\Big<\Big<  \bar p_1|   \bar p_1 \Big>\Big> = \frac{2\kappa^2}{(Q-1)^2}
\nn \\
\Big<\Big< p_1\bar p_1| p_1 \bar p_1 \Big>\Big>
= \kappa^2\frac{\{q\}^2}{\{t\}^2}
\left(\frac{\frac{\{q\}\{t^2\}}{\{t\}\{qt\}}}{Qt^{-2}-1} +
\frac{\frac{\{t\}\{q^2\}}{\{q\}\{qt\}}}{Qq^2-1}\right)
\ee
\be
\Big<\Big<  \Mac_{[1,1]}\{\bar p\} | \Mac_{[1,1]}\{p\}\Big>\Big> =
\kappa \cdot \frac{\{q\}}{\{t\}^2\{t^2\}}\left(
\frac{1}{qt}\cdot\frac{\{q\}\Big\{\frac{q}{t}\Big\}}{Qt^{-2}-1}
- \frac{t}{q}\cdot\frac{\{qt\}\Big\{\frac{q}{t^2}\Big\}}{Q-1}\right)
\nn \\
\Big<\Big< \Mac_{[1,1]}\{\bar p\}| \Mac_{[2]}\{p\}\Big>\Big> =
-\kappa \cdot   \frac{\{q\}\{q^2\}}{\{t\}\{qt\}}\cdot \frac{1}{Qt^{-2}-1}
\nn \\
\Big<\Big<  \Mac_{[2]}\{\bar p\}| \Mac_{[1,1]}\{p\}\Big>\Big> =
-\kappa\cdot \frac{\{q\}\{q^2\}}{\{t\}\{qt\}}\cdot\frac{1}{Qq^2-1}
\nn \\
\Big<\Big<  \Mac_{[2]}\{\bar p\} | \Mac_{[2]}\{p\}\Big>\Big> =
\kappa \cdot \frac{\{q^2\}}{\{t\}\{qt\}^2}\left(-qt\cdot\frac{\{t\}\Big\{\frac{q}{t}\Big\}}{Qq^2-1}
+ \frac{t}{q}\cdot\frac{\{qt\}\Big\{\frac{q^2}{t}\Big\}}{Q-1}\right)
\ee
However, the other components get very complicated. There is another choice, which provides simply looking expressions for all matrix elements. Now we describe it.

\paragraph{Version B:}

In fact, requiring the orthogonality of the generalized Macdonald polynomials, one obtains that the scalar product is defined up to arbitrary 3 components, say, $G_{[2],\emptyset,[2]\big|\emptyset}$, $G_{[1,1],\emptyset\big|[1,1],\emptyset}$, $G_{[1],[1]\big|[1],[1]}$ (note that the components $G_{\emptyset,[2]\big|\emptyset,[2]}$, $G_{\emptyset,[1,1]\big|\emptyset,[1,1]}$ always remain unfixed and are not taken into account). The best choice looks as follows
\begin{eqnarray}
G_{[2],\emptyset\big|[2],\emptyset}&=&\Big(Q-{1\over q^2}\Big)Y,\nn\\
G_{[1,1],\emptyset\big|[1,1],\emptyset}&=&-{\{q^2/t\}\{qt\}^2\over
\{t^2/q\}\{t^2\}\{q^2\}}{(Q-t^2)Y\over q^3t^3}\nn\\
G_{[2],\emptyset\big|[1,1],\emptyset}&=&0\nn\\
G_{[1],[1]\big|[1,1],\emptyset}&=&- {\{q/t\}\{q^2/t\}\{qt\}\{q\}\over
\{q^2\}\{t\}\{t^2/q\}} {Y\over q^4}\nn\\
G_{[1],[1]\big|[2],\emptyset}&=&\{q/t\}{tY\over q^3}\nn\\
G_{[1],[1]\big|[1],[1]}&= & -{\{qt\}\{q^2/t^2\}\{q\}\over
\{t\}\{q^2\}\{t^2/q\}}{(Q-1)t^4Y\over q^7}+{\{qt\}\{q^2/t^2\}\{q\}^2\over
\{q^2\}\{t^2/q\}}{t^3Y\over q^6}+{\{q^2/t\}\{qt\}\{q/t\}\{q\}\over
\{t\}\{q^2\}}{t^2Y\over q^5(Q-1)}\nn\\
G_{\emptyset,[1,1]\big|[1,1],\emptyset}&= &-{\{q^2/t\}\{qt\}\{q/t\}\over
\{t^2/q\}\{q^2\}}{Y\over q^5t}
+{\{q^2/t\}\{q/t\}\{qt\}^2\{t^2\}\over\{q^2\}}{Y\over q^5(Q-1)}\nn\\
G_{\emptyset,[2]\big|[2],\emptyset}&= & {\{q/t\}\{q^2\}\over\{qt\}}{t^2Y
\over q^2}+\{q^2/t\}\{q/t\}{t^2Y\over q^3(Q-1)}\nn\\
G_{\emptyset,[1,1]\big|[2],\emptyset}&= &-\{q/t\}{tY\over q^3}\nn\\
G_{\emptyset,[2]\big|[1,1],\emptyset}&= &{\{q^2/t\}\{q/t\}\over\{t^2/q\}}{Y\over q^4}\nn\\
G_{\emptyset,[1,1]\big|[1],[1]}&= &-{\{q^2/t^2\}\{qt\}\{q/t\}\{q\}\over
\{t\}\{t^2/q\}\{q^2\}}{t^3Y\over q^8}\nn\\
G_{\emptyset,[2]\big|[1],[1]}&= &-{\{q^2/t^2\}\{q/t\}\over\{t^2/q\}}{t^5Y\over q^6}\nn\\
G_{\emptyset,[2]\big|\emptyset,[1,1]}&= &-\{q^2/t\}\{q/t\}^2{t^3Y\over q^6(Q-1)}
\end{eqnarray}
Here we leave $Y$ arbitrary, since the scalar product can be arbitrarily rescaled;
one may require that $G_{[1],[1]\big|[1],[1]}= {\{q\}^2\over\{t\}^2}$ as $Q\to\infty$,
which unambiguously fixes $Y$. Note that an additional pole at $Q=1$ in the denominator emerge iff the second and the third representations coincide. This pole probably comes from the level 1 scalar product. The structure of poles is quite clear from these expressions. The only concern is the element
$G_{\emptyset,[2]\big|\emptyset,[1,1]}$.

This search can be continued to higher levels, however, the calculations becomes much more involved, and the final answer is not available at the moment.

\section{Understanding GMP Hamiltonians\label{pertan}}

We discuss now a specific property of the Hamiltonian (\ref{GEMham}) that makes its eigenfunctions much simpler, in particular, it leads to a cancellation of many denominators that are generally expected to emerge. One can say that the Hamiltonian is actually distinguished by this property.

Our sample example is
\be
\GMac_{[1,1],[1,1]}\{p,\bar p\} = \Mac_{[1,1]}\{p\}\cdot\Mac_{[1,1]}\{\bar p\}
- \frac{t}{q}\Big\{\frac{q}{t}\Big\}\left(
\frac{\Mac_{[2,1]}\{p\}\cdot \Mac_{[1]}\{\bar p\}}{Qq^2t^2-1}
+ \frac{\{q\}\{t^3\}}{\{t\}\{qt^2\}}\cdot
\frac{\Mac_{[1,1,1]}\{p\}\cdot \Mac_{[1]}\{\bar p\}}{Qt^{-2}-1}
\right) + \nn \\
+ \frac{t^3}{q^3}\cdot\frac{\Big\{\frac{q}{t}\Big\}}{Qq^2t^2-1}\cdot \Mac_{[3,1]}\{p\}
+ \frac{\Big\{\frac{q}{t}\Big\}}{\{t\}\{qt\}}\left(
\frac{t^2}{q^2}\cdot\frac{\{qt\}\Big\{\frac{q}{t^2}\Big\}}{Qq^2-1}-
\frac{t^4}{q^2}\cdot\frac{\{q\}\Big\{\frac{q}{t}\Big\}}{Qq^2t^2-1}\right)\cdot\Mac_{[2,2]}\{p\}
+\nn \\
+ \frac{\{q\}\Big\{\frac{q}{t}\Big\}^2}{\{t\}\{qt\}\{qt^2\}\{q^2t^2\}}\left(
\frac{t}{q^3}\cdot\frac{\{t^3\}\{q^2t^2\}}{Qt^{-2}-1}
- \frac{t^5}{q}\cdot\frac{\{t^2\}\{q^2t\}}{Qq^2t^2-1}\right)\cdot \Mac_{[2,1,1]}\{p\}
+ \nn \\
+\frac{\{q\}\{t^3\}\{t^4\}\Big\{\frac{q}{t}\Big\}}{\{t\}^2\{t^2\}\{qt^2\}\{qt^3\}}
\left(\frac{t^2}{q^2}\cdot\frac{\{qt\}\Big\{\frac{q}{t^2}\Big\}}{Qt^{-4}-1}
- \frac{t^4}{q^2}\cdot\frac{\{q\}\Big\{\frac{q}{t}\Big\}}{Qt^{-2}-1}
\right)\cdot\Mac_{[1,1,1,1]}\{p\} \ \ \ \ \ \
\label{GEM1111}
\ee
which we are going to obtain as an eigenfunction of the Hamiltonian (\ref{GEMham}) by perturbation theory.

Let us redefine the Hamiltonian (\ref{GEMham}) by introducing parameters $\varepsilon_n$, which are the perturbation parameters:
\be
\hat {\cal H}_\varepsilon   =
\frac{1}{t^2-1}\left\{
\oint \frac{dz}{z}
\exp\left(\sum_n \frac{(1-t^{-2n})p_nz^n}{n}\right)
\exp\left(\sum_n \frac{q^{2n}-1}{z^n}\, \frac{\p}{\p p_n}\right)
-1\right\} + \ \ \ \ \ \ \
\nn \\
+ \frac{1}{Q}\cdot \frac{1}{t^2-1}\left\{
\oint \frac{dz}{z}\exp\left(\sum_n \frac{(1-t^{-2n})z^n}{n}
\Big(\underline{\varepsilon_n \left(1-(t/q)^{2n}\right)p_n}+\bar p_n)\right)
\exp\left(\sum_n \frac{q^{2n}-1}{z^n}\, \frac{\p}{\p \bar p_n}\right)
-1\right\}
\label{GEMhame}
\ee
Then (\ref{GEM1111}) gets $\varepsilon$-deformed into
$$
\GMac_{[1,1],[1,1]}^\varepsilon\{p,\bar p\} = \Mac_{[1,1]}\{p\}\cdot\Mac_{[1,1]}\{\bar p\}
- \varepsilon_1\cdot\frac{t}{q}\Big\{\frac{q}{t}\Big\}\left(
\frac{\Mac_{[2,1]}\{p\}\cdot \Mac_{[1]}\{\bar p\}}{Qq^2t^2-1}
+ \frac{\{q\}\{t^3\}}{\{t\}\{qt^2\}}\cdot
\frac{\Mac_{[1,1,1]}\{p\}\cdot \Mac_{[1]}\{\bar p\}}{Qt^{-2}-1}
\right) + 
$$
$$
+ \Big\{\frac{q}{t}\Big\} \left(\frac{t^3}{q^3}\cdot\frac{\varepsilon_1^2}{Qq^2t^2-1}
+ \frac{t(t^2+1)(q^2+t^2)}{2q^3}\cdot
 \boxed{\frac{\varepsilon_2-\varepsilon_1^2}{ Qq^4t^2+Qq^2t^2-t^2-1}}\,\right)
\cdot \Mac_{[3,1]}\{p\} +
$$
$$
+ \frac{\Big\{\frac{q}{t}\Big\}}{\{t\}\{qt\}}\left(
\varepsilon_1^2\cdot\frac{t^2}{q^2}\cdot\left(\frac{\{qt\}\Big\{\frac{q}{t^2}\Big\}}{Qq^2-1}
- \frac{t^2\{q\}\Big\{\frac{q}{t}\Big\}}{Qq^2t^2-1}\right)
- \frac{t\{t\}^2(q^2+1)(q^2+t^2)}{2q^4}\cdot\frac{\varepsilon_2-\varepsilon_1^2 }{Qq^2-1}
\right)
\cdot\Mac_{[2,2]}\{p\}
+
$$
$$
\!\!\!\!\!\!\!\!\!\!\!\!\!\!\!\!\!\!\!\!\!\!
+ \left(
 \frac{\varepsilon_1^2\cdot\{q\}\Big\{\frac{q}{t}\Big\}^2}{\{t\}\{qt\}\{qt^2\}\{q^2t^2\}}\left(
\frac{t}{q^3}\cdot\frac{\{t^3\}\{q^2t^2\}}{Qt^{-2}-1}
- \frac{t^5}{q}\cdot\frac{\{t^2\}\{q^2t\}}{Qq^2t^2-1}\right)
+\frac{t^3}{q^3}\cdot\frac{\{q\}^2\Big\{\frac{q}{t}\Big\}^2(t^2+1)(q^2+t^2)}{2\{qt\}\{q^2t^2\}}
\cdot\boxed{\frac{\varepsilon_2-\varepsilon_1^2}{Qq^2t^4+Q-t^4-t^2}}\,\right)
\cdot \Mac_{[2,1,1]}\{p\}
+ 
$$
$$
+\frac{\{q\}\{t^3\}\{t^4\}\Big\{\frac{q}{t}\Big\}}{\{t\}^2\{t^2\}\{qt^2\}\{qt^3\}}
\left(\varepsilon_1^2\cdot \frac{t^2}{q^2}\cdot\left(\frac{\{qt\}\Big\{\frac{q}{t^2}\Big\}}{Qt^{-4}-1}
-  \frac{t^2\{q\}\Big\{\frac{q}{t}\Big\}}{Qt^{-2}-1}\right)
- \frac{t\{t\}^2(q^2+1)(q^2+t^2)}{2q^4}\cdot\frac{\varepsilon_2-\varepsilon_1^2}{Qt^{-4}-1}
\right)\cdot\Mac_{[1,1,1,1]}\{p\} \ \ \ \ \ \
$$
Clearly, the items with ``bad" denominators (boxed) vanish when $\varepsilon_2=\varepsilon_1^2$,
i.e. this is a special cancellation associated with a peculiar choice among
triangular(!) Hamiltonians.
This simple observation raises a bunch of important questions:

\begin{itemize}
\item{
What kind of symmetry this is, so that the things work this way
in all orders of perturbation theory?
}

\item{
Why cancellation of bad denominators (poles) in the GMP guarantees that they are canceled
also in the scalar product?
}

\item{
Is this the same symmetry that allows one to easily adjust all the higher Hamiltonians,
i.e. how is it related to integrability?
}
\end{itemize}

\section{Generalized Schur functions}

At  $q=t$, the generalized Macdonald polynomials trivialize:
\be
\left.\GMac_{R,\bar R}\{p,\bar p\}\right|_{q=t} = \Schur_{R}\{p\}\Schur_{\bar R}\{\bar p\}
\ee
and the $Q$-dependence remains only in the eigenvalues
\be
\lambda_{R,\bar R}= \lambda_R+Q^{-1}\lambda_{\bar R} =
\sum_{i=1}^{l_R} \frac{q^{2r_i}-1}{q^{2i}}
+ \frac{1}{Q}\sum_{i=1}^{l_{R'}} \frac{q^{2r'_i}-1}{q^{2i}}
\ee
This is because the deformation of GMP is proportional to $\{q/t\}$.

There is, however, another interesting possibility: consider again the Hamiltonian (\ref{GEMham}), however, this time without specifying the parameters $\epsilon_n$:
\be
\hat {H}_\epsilon   =
\frac{1}{t^2-1}\left\{
\oint \frac{dz}{z}
\exp\left(\sum_n \frac{(1-t^{-2n})p_nz^n}{n}\right)
\exp\left(\sum_n \frac{q^{2n}-1}{z^n}\, \frac{\p}{\p p_n}\right)
-1\right\} + \ \ \ \ \ \ \
\nn \\
+ \frac{1}{Q}\cdot \frac{1}{t^2-1}\left\{
\oint \frac{dz}{z}\exp\left(\sum_n \frac{(1-t^{-2n})z^n}{n}
\Big(\underline{\epsilon_n
p_n}+\bar p_n)\right)
\exp\left(\sum_n \frac{q^{2n}-1}{z^n}\, \frac{\p}{\p \bar p_n}\right)
-1\right\}
\label{Schurhame}
\ee
Its first eigenfunctions
are:
\be
\begin{array}{cl}
{\GMac}^\epsilon_{[1],\emptyset}\{p,\bar p\} = &
\Schur_{[1]}\{p\}
 \\
{\GMac}^\epsilon_{\emptyset,[1] }\{p,\bar p\} = &
\Schur_{[1]}\{\bar p\} - \frac{\epsilon_1}{Q-1}\Schur_{[1]}\{p\}\\
&\\
{\GMac}^\epsilon_{[1,1],\emptyset}\{p,\bar p\} =&\Schur_{[1,1]}\{p\} =\Mac_{[1,1]}\{p\}
  \\
{\GMac}^\epsilon_{[2],\emptyset}\{p,\bar p\} =
&\Schur_{[2]}\{p\}-{\frac{{\left\{\frac{q}{t}\right\}}}{\{qt\}}\Schur_{[1,1]}\{p\} }
= \Mac_{[2]}\{p\}
  \\
{\GMac}^\epsilon_{[1],[1]}\{p,\bar p\} =&
\Schur_{[1]}\{p\}\Schur_{[1]}\{\bar p\} - \frac{\epsilon_1}{Qq^2-1} \Schur_{[2]}\{p\}
+\epsilon_1\left({\frac{ \left\{\frac{q}{t}\right\}}{\{qt\}}\frac{1}{Qq^2-1}}
- \frac{ \{q\}\{t^2\}}{\{t\}\{qt\}}\frac{1}{Qt^{-2}-1}\right)\Schur_{[1,1]}\{p\}
\\
&=\Mac_{[1]}\{p\}\Mac_{[1]}\{\bar p\} - \frac{\epsilon_1}{Qq^2-1} \Mac_{[2]}\{p\}
- \frac{\epsilon_1\cdot \{q\}\{t^2\}}{\{t\}\{qt\}(Qt^{-2}-1)} \Mac_{[1,1]}\{p\}
\\ &\\
{\GMac}^\epsilon_{\emptyset,[1,1]}\{p,\bar p\} = &
\Schur_{[1,1]}\{\bar p\}-\frac{\epsilon_1}{Qt^2-1}\Schur_{[1]}\{p\}\Schur_{[1]}\{\bar p\}
+ \underline{\left(\frac{\epsilon_1^2\cdot t}{q\left\{\frac{q}{t}\right\} (Qt^2-1)}
+\frac{\{t^2\}}{2\{t\}\left\{\frac{q}{t}\right\} }
\boxed{\frac{(q^2-t^2)\epsilon_2-(q^2+t^2)\epsilon_1^2}{Qt^2(q^2+1)-(t^2+1)}}\right)}
\Schur_{[2]}\{p\}
-  \\
&-\left(\frac{t\cdot\epsilon_1^2}{\{t\}(Qt^2-1)}
+ \frac{(q^2+1)t^2\{t\}^2\epsilon_2 - q\{q\}(t^4+1)\epsilon_1^2}{2qt^2\{t\}\{qt\}(Q-1)}
+  \frac{\{t^2\}}{2q\{t\}\{qt\}}\boxed{
\frac{(q^2-t^2)\epsilon_2-(q^2+t^2)\epsilon_1^2}{Qt^2(q^2+1)-(t^2+1)}}
\right)\Schur_{[1,1]}\{p\}
  \\
{\GMac}^\epsilon_{\emptyset,[2]}\{p,\bar p\} =
& \Schur_{[2]}\{\bar p\}
-  {\frac{\left\{\frac{q}{t}\right\}}{\{qt\}}\Schur_{[1,1]}\{\bar p\}}
 - \frac{\epsilon_1}{Qq^{-2}-1}\frac{\{t\}\{q^2\}}{\{q\}\{qt\}}
\, \Schur_{[1]}\{p\}\Schur_{[1]}\{\bar p\}
 +
 \\
& \ \ \ \ \ \ \ \ \ \ \ \ \ \ \ \ \ \ \ \ \ \
+\left(\frac{\epsilon_1^2\{t\}}{q^2\{q\}\{qt\}(Qq^{-2}-1)}
 - \frac{q^2\{q\}^2(t^2+1)\epsilon_2-(q^4+1)t\{t\}\epsilon_1^2}{2q^2t\{q\}\{qt\}(Q-1)}
\right) \Schur_{[2]}\{p\}
+ \\
&
\!\!\!\!\!\!\!\!\!\!\!\!\!\!\!\!\!\!\!\!\!\!\!\!\!\!\!
\!\!\!\!\!\!\!\!\!\!\!\!\!\!\!\!\!
+ \left( \frac{\left\{\frac{q}{t}\right\}
\left(q^2\{q\}^2(t^2+1)\epsilon_2+t\{t\}(q^4+1)\epsilon_1^2\right)}{2q^2t\{q\}\{qt\}^2(Q-1)}
+\underline{
\frac{\epsilon_1^2\{t\}(q^4t^2+q^4-q^2t^2-t^2)}
{q^3t\{q\}\{qt\}\left\{\frac{q}{t}\right\}(Qq^{-2}-1)}
+ \frac{t\{q^2\}^2\{t^2\}}{2\{q\}\{qt\}^2\left\{\frac{q}{t}\right\}}\boxed{
\frac{(q^2-t^2)\epsilon_2-(q^2+t^2)\epsilon_1^2}{Q(t^2+1)-t^2(q^2+1)}}
}\right)  \Schur_{[1,1]}\{p\}
  \\ & \\
\ldots
\end{array}\nn
\ee
As we discussed in the previous section, the bad poles (which are boxed) disappear at $\epsilon_2= \frac{q^2+t^2}{q^2-t^2}\cdot\epsilon_1^2$,
i.e. when $(1-t^4/q^4)\epsilon_2 = (1+t^2/q^2)^2\epsilon_1^2$.

Now we consider the $t=q$ case {\it without imposing any restrictions on $\epsilon_n$}.
Note that the underlined terms with bad poles are singular in the limit $q=t$.
The singularity is resolved by application of the l'Hopital rule,
and, in result, the bad poles become ``good" {\it double} poles (underlined):
\be
{\cal S}^\epsilon_{[1],\emptyset}\{p,\bar p\} = &
\Schur_{[1]}\{p\}
\nn \\
{\cal S}^\epsilon_{\emptyset,[1] }\{p,\bar p\} = &
\Schur_{[1]}\{\bar p\} - \frac{\epsilon_1}{Q-1}\Schur_{[1]}\{p\}
\nn \\ \nn\\
{\cal S}^\epsilon_{[1,1],\emptyset}\{p,\bar p\} =&\Schur_{[1,1]}\{p\}
\nn  \\
{\cal S}^\epsilon_{[2],\emptyset}\{p,\bar p\} =
&\Schur_{[2]}\{p\}
\nn  \\
{\cal S}^\epsilon_{[1],[1]}\{p,\bar p\} =&
\Schur_{[1]}\{p\}\Schur_{[1]}\{\bar p\} - \frac{\epsilon_1}{Qq^2-1} \Schur_{[2]}\{p\}
-  \frac{\epsilon_1}{Qq^{-2}-1} \Schur_{[1,1]}\{p\}
\nn\\
{\cal S}^\epsilon_{\emptyset,[1,1]}\{p,\bar p\} = &
\Schur_{[1,1]}\{\bar p\}-\frac{\epsilon_1}{Qq^2-1}\Schur_{[1]}\{p\}\Schur_{[1]}\{\bar p\}
+\left(\underline{\frac{Qq^4\epsilon_1^2}{(q^2+1)(Qq^2-1)^2}}
+ \frac{\epsilon_2-\epsilon_1^2}{2(Qq^2-1)}\right)\Schur_{[2]}\{p\} -
\nn\\
& -\left(
\frac{\epsilon_1^2}{q^4-1}\left(\frac{q^4}{Qq^2-1}-\frac{1}{Q-1}\right)
+ \frac{\epsilon_2-\epsilon_1^2}{2(Q-1)}\right)\Schur_{[1,1]}\{p\}
\nn \\
{\cal S}^\epsilon_{\emptyset,[2]}\{p,\bar p\} =
& \Schur_{[2]}\{\bar p\} - \frac{\epsilon_1}{Qq^{-2}-1}\Schur_{[1]}\{p\}\Schur_{[1]}\{\bar p\}
+\left(\frac{\epsilon_1^2}{\{q^2\}}\Big(\frac{q^{-2}}{Qq^{-2}-1}-\frac{q^2}{Q-1}\Big)
- \frac{\epsilon_2-\epsilon_1^2}{2(Q-1)}
\right) \Schur_{[2]}\{p\} +
\nn \\
&+ \left(\underline{\frac{Q\epsilon_1^2}{q^2(q^2+1)(Qq^{-2}-1)^2}}
+\frac{\epsilon_2-\epsilon_1^2}{2(Qq^{-2}-1)}\right)
  \Schur_{[1,1]}\{p\}
  \nn \\
\ldots
\ee
They becomes a product of two Schur functions upon putting $\epsilon_n=0$, which is the case at $t=q$ for $\epsilon_n=1-q^{2n}/t^{2n}$.

Thus, we define non-trivial generalized Schur polynomials (GSP), which are a {\it simpler} version of GMP, but {\it not} their particular case unless one takes the limit (\ref{Minf}).
Looking at GSP can help to reveal and describe the essential properties of
generic generalized polynomials, and this can be the first step
to untie them from peculiarities of the Macdonald polynomials.

The Hamiltonian (\ref{Schurhame}) at $t=q$ unambiguously defines the GSP's, however, higher Hamiltonians are no longer given by the construction of s.3.2.

\section{Cauchy formula vs triangular expansion\label{sCauchy}}

\paragraph{The Cauchy formula.}
The Cauchy formula is always correct and is a direct corollary of the orthogonality relation. In this Appendix, we demonstrate how it works for the GMP. We start, however, with the usual Macdonald polynomials. Then, the Cauchy formula has the form
\be
\sum_R {\Mac_R\{p\}\Mac_R\{p'\}\over ||\Mac_R||^2}=\exp\left(\sum_k{\{t^k\}\over\{q^k\}}{p_kp'_k\over k}\right)
\ee
In order to prove this formula, one has to act with the operator $\Mac_Q\left\{k{\{q^k\}\over\{t^k\}}{\partial\over\partial p_k}\right\}$ on the both sides of this formula and then put  all $p_k=0$. Then, the r.h.s. is trivially equal to $\Mac_Q\{p'\}$, while, at the l.h.s., one has to use the orthogonality condition (\ref{orc}) and realize the scalar product (\ref{sp}) by the differential operation
\be
\Big< { p}_{\Delta}\Big| F\{p\} \Big>_{(q,t)}=z_\Delta\cdot
\left(\prod_{i=1}^{l_\Delta} {\{q^{\delta_i}\}\over\{t^{\delta_i}\}}{\partial\over\partial p_{\delta_i}}\right)F\{p\}\Biggr|_{p_k=0}
\ee
in order to obtain $\Mac_Q\{p'\}$ too. Now one can use the transposition formula for the Macdonald polynomials
\be\label{trM}
\Mac_R\{p_k\}=||\Mac_R||^2\cdot\Biggr[\Mac_{R^\vee}\Big\{(-1)^{k+1}{\{q^k\}\over\{t^k\}}p_k\Big\}\Biggr]_{q\leftrightarrow t}
\ee
in order to obtain
\be\label{MC}
\sum_R (-1)^{|R|}\Mac_R\{p\}\Big[\Mac_{R^\vee}\{p'\}\Big]_{q\leftrightarrow t}=\exp\left(-\sum_k{p_kp'_k\over k}\right)
\ee
The superscript $\vee$ here denotes the transposition of the Young diagram.

This consideration demonstrates that, in order to obtain the Cauchy formula for the GMP's one has to use both $\GMac_{R_1,R_2}$ and $\widetilde\GMac_{R_1,R_2}$, due to the orthogonality relation (\ref{orthodualGMP}). Then, the Cauchy formula acquires the form
\be
\sum_{R_1,R_2} {\widetilde\GMac_{R_1,R_2}\{p,\bar p\}
\GMac_{R_1,R_2}\{p',\bar p'\}\over ||\Mac_{R_1}||^2||\Mac_{R_2}||^2}=\exp\left(\sum_k{\{t^k\}\over\{q^k\}}{p_kp'_k+\bar p_k\bar p_k'\over k}\right)
\ee
The transposition formula now looks like
\be
\GMac_{R_1,R_2}\{p_k\}=||\Mac_{R_1}||^2||\Mac_{R_2}||^2
\cdot\Biggr[\GMac_{R_1^\vee,R_2^\vee}\Big\{(-1)^{k+1}{\{q^k\}\over\{t^k\}}p_k,(-1)^{k+1}{\{q^k\}\over\{t^k\}}\bar p_k
\Big\}\Biggr]_{q\leftrightarrow t}
\ee
and the Cauchy formula is
\be
\sum_{R_1,R_2} (-1)^{|R_1|+|R_2|}\widetilde\GMac_{R_1,R_2}\{p,\bar p\}
\GMac_{R_1^\vee,R_2^\vee}\{p',\bar p'\}=\exp\left(-\sum_k{p_kp'_k+\bar p_k\bar p_k'\over k}\right)
\ee

\paragraph{The triangular structure.}
Let us now note that the Cauchy formula for the Schur polynomials
\be
\sum_R (-1)^{|R|}\Schur_R\{p\} \Schur_{R^\vee}\{p'\} = \exp\left(-\sum_k \frac{p_kp'_k}{k}\right)
\ee
is immediately equivalent to
\be
\sum_R (-1)^{|R|}\Mac_R\{q,t,p\} \Mac_{R^\vee}\{t,q, p'\} = \exp\left(-\sum_k \frac{p_kp'_k}{k}\right)
\ee
because of the identity
\be\label{KK}
\sum_R{\cal K}_{R'R}(q,t) {\cal K}_{{R''}^\vee,R^\vee}(t,q) = \delta_{R'R''}
\ee
valid for the triangular transformation (Macdonald-Kostka) matrix
\be
\Mac_R\{q,t,p\} =\sum_Q{\cal K}_{R,Q}(q,t)\Schur_Q\{p\}
\ee
(which is straightforwardly lifted to Kerov-Kostka matrix with arbitrary Kerov couplings $g_k$).
In order to prove eq.(\ref{KK}), note that the scalar product (\ref{sp}) of the Schur functions is equal to
\be
\mu_{R_1,R_2}(q,t)=\Big<\Schur_{R_1}\Big|\Schur_{R_2}\Big>_{(q,t)}=\sum_{R}\Big({\cal K}^{-1}\Big)_{R_1,R}(q,t)
\cdot ||\Mac_R(q,t)||^2\cdot\Big({\cal K}^{-1}\Big)_{R_2,R}(q,t)
=\sum_\Delta \ {\psi_{R_1}(\Delta)\psi_{R_2}(\Delta)\over z_\Delta}\cdot g_\Delta\nn
\label{Schurscaprog1}
\ee
where
\be
g_\Delta:=\left(\prod_{i=1}^{l_\Delta} {\{q^{\delta_i}\}\over\{t^{\delta_i}\}}\right)
\ee
Thus, the inverse Macdonald-Kostka matrix diagonalizes the matrix $\mu_{R,R'}$ with the diagonal elements being $||\Mac_R||^2$. We write it in the matrix form
\be
{\bf \mu}={\bf K}^{-1}\cdot {\bf M}\cdot \widetilde {\bf K}^{-1}
\ee
where ${\bf M}$ is the diagonal matrix with elements $||\Mac_R(q,t)||^2$ and the tilde denotes the matrix transposition.

At the same time, the inverse matrix
\be
\Big(\mu^{-1}\Big)_{R_1,R_2}(q,t)=\sum_\Delta \ {\psi_{R_1}(\Delta)\psi_{R_2}(\Delta)\over z_\Delta}\cdot {1\over g_\Delta}=
\mu_{R_1,R_2}(t,q)
\label{Schurscaprog}
\ee
is diagonalized by the Macdonald-Kostka matrix itself:
\be\label{muin}
{\bf \mu}^{-1}={\bf K}\cdot {\bf M}^{-1}\cdot \widetilde {\bf K}=\widetilde{\bf K}\cdot {\bf M}^{-1}\cdot {\bf K}
\ee
Now it remains to notice that
\be
 ||\Mac_R(q,t)||^2= ||\Mac_{R^\vee}(t,q)||^{-2}
\ee
Let us denote the interchanging $q$ and $t$ in a matrix with a bar and introduce the matrix $\Lambda:=\delta_{R,R^\vee}$
so that this latter identity reads as
\be\label{ls}
{\bf M}=\Lambda\cdot\overline{\bf M}^{-1}\cdot\Lambda
\ee
Then one obtains from (\ref{Schurscaprog})-(\ref{ls})
\be\label{KK2}
\mu^{-1}=\bar\mu\ \ \ \ \ \ \Longrightarrow\ \ \ \ \ \ {\bf K}\cdot\Lambda\cdot\widetilde{\overline{\bf K}}\cdot\Lambda=1
\ee
which is exactly (\ref{KK}).

This consideration directly extended to the case of GMP and of more general Kerov functions. To better understand the
the reason for additional matrices $\Lambda$ emerging in (\ref{KK2}) and somewhat strangely looking formula (\ref{KK}),
it is instructive to look at a straightforward generalization of the Macdonald polynomials, Kerov functions \cite{Kerov,MMkerov}.

\paragraph{Kerov functions.}
One can consider an extension of the scalar product (\ref{sp}) to
\be
\Big< { p}_{\Delta}\Big| { p}_{\Delta'} \Big>_{(g)} =
z_\Delta \cdot \delta_{\Delta,\Delta'}\cdot
\left(\prod_{i=1}^{l_\Delta} g_{\delta_i}\right)
\label{scapro}
\ee
which induces the Kerov functions\footnote{The Kerov {\it function} should not be confused
with much better known  ``Kerov character polynomials",
also associated with the name of S.Kerov \cite{Kerpols}} \cite{Kerov,FZ,Kerovmore,MMkerov}, which are defined to be {\it triangular} transforms from the Schur functions orthogonal with the scalar product (\ref{scapro}). The Young diagrams are again ordered here lexicographically.
However, in this case, there are two systems of symmetric functions, since the Kostka-Kerov matrices ${\cal K}_{{R,R'}}^{(g)}$ between diagrams placed differently in two different orderings are non-zero:
\be
\begin{array}{c}
\SchurP^{(g)}_R\{ p\} =  {\Schur}_{R}\{p\}
+ \sum_{{R'< R}} {\cal K}_{{R,R'}}^{(g)} \cdot {\Schur}_{R'}\{p\}
\\ \\
\widehat{\SchurP}_R^{(g )}\{p\}
=\Schur_{R}\{p\}\ + \  \sum_{ R'^\vee> R^\vee}
\widehat{\cal K}^{(g )}_{R'^\vee,R^\vee}\cdot\Schur_{R'}\{p\}
\end{array}
\label{PthroughSchurs1}
\ee
 Since the orderings begin to differ at level 6, the two system of the Kerov functions become different at level 6.

The Cauchy formula in this case looks like
\be
\sum_{R}(-)^{|R|}\SchurP_R^{(g)}\{p\}\cdot\widehat{\SchurP}^{(g^{-1})}_{R^\vee}\{-{p'}\}
=\sum_R \frac{\SchurP_R^{(g)}\{p \}\cdot\SchurP_R^{(g)}\{p'^\wedge\}}{||\SchurP_R^{(g)}||^2}
= \exp\left( \sum_k \frac{p_kp_{k}'}{k}\right)
\label{Cauchy}
\ee
since the rule that relates the Kerov functions of transposed Young diagrams (a counterpart of (\ref{trM})) is
\be
\SchurP_{R }^{(g)}\{p\} =
(-)^{|R|}\cdot||\SchurP_{R }^{(g)}||^2 \cdot
\widehat{\SchurP}^{(g^{-1})}_{R^\vee}\left\{ -{p_k\over g_k}  \right\}
\ee
The identity (\ref{KK}) becomes, in the Kerov case,
\be\label{KK3}
\sum_R{\cal K}^{(g )}_{R'R}(q,t) \widehat{\cal K}^{(g )}_{R,R''}(t,q) = \delta_{R'R''}
\ee
Notice that the definition of $\widehat{\cal K}$ in (\ref{PthroughSchurs1}) differs from ${\cal K}$. In the Macdonald case, $\widehat{\cal K}$ is related with ${\cal K}$ by conjugation with the matrix $\Lambda$, which explains emergence of this latter in (\ref{KK2}).

Formula (\ref{KK3}) is proved in the same way as in s.\ref{sCauchy} with taking into account that
\be
 ||\SchurP^{(g)}_R||^2= ||\widehat\SchurP^{(g^{-1})}_{R^\vee}||^{-2}
\ee

\section{Conclusion}

The goal of this paper was to understand the notion of {\it generalized}
symmetric polynomials.
In the literature they are actually known in just one particular example of the
generalized Macdonald polynomials, which appear in the theory of  AGT relations,
where  $q,t$-deformed Selberg-Kadell integrals of GMP
presumably reproduce Nekrasov functions.
The question, however, is to find more algebraic and, most important,
generalizable definitions.

Usually, one uses two approaches, which we described and discussed in the text.
Both are based on the  triangularity property, which is now
understood to be the basic one for the {\it actual} theory of symmetric functions
and their application to integrability in general and supersymmetric gauge theories
in particular.
All the relevant special functions in these fields are triangular combinations
of Schur functions: triangular with respect to one or another kind of
lexicographical ordering of Young diagrams.
However, the triangularity {\it per se} is not enough.
The two approaches consist of imposing orthogonality conditions either directly
or through a requirement of forming the eigenfunctions of Hermitian Hamiltonians.

We discussed advantages and drawbacks of these two approaches
in application to GMP.

The problem with straightforward orthogonality is the complexity (and ambiguity)
of the relevant scalar product, which we did not even manage to construct
in the full form.
For naive scalar products, the GMP's are not orthogonal: one needs to introduce
a complementary set of ``dual" GMP, and then the question is to impose relations
between the dual and original GMP.

The Hamiltonian approach in the case of GMP is more successful, and we advanced it
further to explicit description of higher Hamiltonians.
However, even the triangularity property is somewhat obscure in this case,
even for the {\it ordinary} Macdonald functions, and this makes the construction
of Hamiltonians and the very fact of their existence a kind of mysterious art,
preventing making integrability from becoming a clear deductive piece of science.
This is a serious drawback, if one looks for generalizations,
which are desperately needed,
because most partition functions in quantum field theory are related
to still unknown integrable systems.
Moreover, even in pure mathematics, Hamiltonians are not yet known
for an immediate generalization from the Macdonald polynomials to the Kerov functions.

We hope that our detailed presentation of these issues, of many problems
which remain unsolved, and of many ``miracles" which can provide the clues
for their resolution, will attract an attention and help to advance the subject
in the near future.

\section*{Acknowledgements}

Our work is supported in part by the grant of the Foundation for the Advancement of Theoretical Physics ``BASIS" (A.Mir., A.Mor.), by  RFBR grants 19-01-00680 (A.Mir.) and 19-02-00815 (A.Mor.), by joint grants 19-51-53014-GFEN-a (A.Mir., A.Mor.), 19-51-50008-YaF-a (A.Mir.), 18-51-05015-Arm-a (A.Mir., A.Mor.), 18-51-45010-IND-a (A.Mir., A.Mor.).

\end{document}